\documentclass[ALICE,manyauthors]{cernphprep}
\usepackage[comma,square,numbers,sort&compress]{natbib}
\usepackage{hyperref}
\usepackage{lineno}
\usepackage{xspace}
\usepackage[section]{placeins}
\usepackage[dvipsnames]{xcolor}
\usepackage{multirow}

\usepackage[T1]{fontenc}
\usepackage{orcidlink}

\begin{document}

%%%%%%%%%%%%%%%  Title page %%%%%%%%%%%%%%%%%%%%%%%%
\begin{titlepage}
% the dates below correspond to CERN approval
% please don't touch: EB chairs will take care
\PHyear{2025}       % required, will be obtained from CERN
\PHnumber{236}      % required, will be obtained from CERN
\PHdate{13 October}  % required, will be obtained from CERN
%%%%%%%%%%%%%%%%%%%%%%%%%%%%%%%%%%%%%%%%%%%%%%%%%%%%

%%% Put your own title + short title here:
\title{One- and three-dimensional identical charged-kaon femtoscopic correlations in Pb--Pb collisions at
$\sqrt{s_\mathrm{NN}}=5.02$~TeV}
\ShortTitle{1D and 3D K$^\pm$K$^\pm$ femtoscopic correlations in Pb--Pb collisions}   % appears on right page headers

%%% Do not change the next lines
\Collaboration{ALICE Collaboration\thanks{See Appendix~\ref{app:collab} for the list of collaboration members}}
\ShortAuthor{ALICE Collaboration} % appears on right page headers, do not change

\begin{abstract}
The identical charged-kaon correlations induced by quantum-statistics effects and final-state interactions are measured in Pb--Pb collisions at \mbox{$\sqrt{s_{\rm NN}} = 5.02$}~TeV. The results of one- (1D) and three-dimensional (3D) analyses show that the obtained system-size parameters (radii) are smaller for more peripheral collisions and decrease with increasing pair transverse momentum $k_{\rm T}$. The 1D parameters agree within uncertainties with those obtained in Pb--Pb collisions at \mbox{$\sqrt{s_{\rm NN}}=2.76$}~TeV. The observed power-law dependence of the extracted 3D radii as a function of the pair transverse momentum is a signature of the collective flow in the particle-emitting system created in Pb--Pb collisions. This dependence is well reproduced by the integrated hydrokinetic model calculations except for the outward projection of the radius (measured in the longitudinally co-moving system) for the most central collisions. The time of maximal emission for kaons is extracted from the 3D analysis in a wide collision centrality range from 0 to 90\%. Its reduction with decreasing charged-particle multiplicity is well reproduced by the hydrokinetic model predictions, and means that kaons are emitted earlier in more peripheral events.

\end{abstract}
\end{titlepage}

\setcounter{page}{2} %please do not remove this line

%%%%%%%%%%%%%%%%%%%%%%%%%%%%%%%%
% begin main text
%%%%%%%%%%%%%%%%%%%%%%%%%%%%%%%%

\section{Introduction}

The collisions of heavy ions at ultrarelativistic energies result in the creation of the deconfined state of strongly interacting matter characterized by partonic degrees of freedom (quark--gluon plasma -- QGP) that was predicted~\cite{SHURYAK198071} and whose signatures were established for the first time at the SPS~\cite{Gazdzicki:1998rx,NA50:2000brc,Lourenco:2001wi} and RHIC~\cite{BRAHMS:2004adc,PHOBOS:2004zne,STAR:2005gfr,PHENIX:2004vcz} accelerators. Studying the QGP is one of the main goals of ALICE (A Large Ion Collider Experiment) at the Large Hadron Collider (LHC) and other experiments investigating heavy-ion collisions  (Section 1.3 in Ref.~\cite{ALICE:2022wpn}).
The system created in such collisions behaves as a strongly coupled liquid with small viscosity~\cite{STAR:2005gfr,PhysRevLett.94.111601, PhysRevC.76.024905} and is well described within models containing a hydrodynamical stage of evolution in the momentum sector in terms of the transverse momentum spectra~\cite{Guillen:2020nul}, the radial and elliptic flow phenomena~\cite{Retiere:2003kf}, and the event-by-event fluctuations~\cite{Qian:2016pau,Werner:2023zvo}. In the space--time domain, its properties can be investigated via correlations between particles close in momentum. This is achieved using the femtoscopy technique~\cite{SINYUKOV1994589,AKKELIN1995525}, which accesses a region of particle emission, characterized by the homogeneity length~\cite{Lisa:2005dd}. Femtoscopy studies particle correlations due to quantum-statistics effects and final-state interactions. It is widely used to study the QGP evolution in heavy-ion collision physics~\cite{Lednicky2004,Pratt:1984su,Lisa:2005dd}.

Femtoscopic correlations due to Bose--Einstein statistics were first observed in the study of pions~\cite{Goldhaber:1960sf}. Since pions are the most abundant particles produced in high-energy collisions, they were investigated more than any other hadrons for femtoscopic correlations. The next most produced particle species in heavy-ion collisions are kaons. The advantage of kaon analyses is that they are expected to offer a cleaner signal compared to pions, because it includes smaller contributions from short-lived resonance decays. Using pion and kaon analyses results, one can compare the effective radii of their emission regions in order to check the phenomenon of the pair transverse mass $m_\mathrm{T}$ scaling~\cite{Makhlin:1987gm,Lisa:2005dd}. It predicts a universal dependence of the radii for various particle types on \mbox{$m_\mathrm{T} = \sqrt{k_\mathrm{T}^2 + m^2}$}, where \mbox{$k_\mathrm{T} = | \mathbf{p}_{\rm T,1} + \mathbf{p}_{\rm T,2} |/2$} is the pair transverse momentum for particles with transverse momenta $\mathbf{p}_{\rm T,1}$ and $\mathbf{p}_{\rm T,2}$ and \mbox{$m=(m_1+m_2)/2$} is the average mass of the correlating particles with masses $m_1$ and $m_2$. As shown in Refs.~\cite{Shapoval:2014wya,STAR:2013che,PHENIX:2015jaj,ALICE:2017iga}, the $m_\mathrm{T}$ scaling expected by models that do not include rescattering processes~\cite{Kisiel:2014upa} is broken for pions and kaons. The K$^*$(892) resonance has a particularly strong effect, as it decays mostly into a kaon and a pion in the hadronic matter ($c\tau \approx 4.0$~fm), leading to the possibility of an interplay with regeneration effects. All these effects are included in the full hydrokinetic model simulations~\cite{SINYUKOV2016227} with the UrQMD cascade~\cite{Knospe:2015nva} as an afterburner. It should also be noted that the $m_{\rm T}$ scaling can be broken without any rescattering effect in the blast-wave model scenarios (see~Ref.~\cite{Retiere:2003kf}). However, they are not discussed in the present article since there are no calculations performed in such models for the collision systems and energies studied in this work. The decrease of the K$^{\pm}$K$^{\pm}$ femtoscopic radii with increasing $m_\mathrm{T}$ ($k_\mathrm{T}$) observed previously by the ALICE Collaboration ~\cite{ALICE:2015hvw,ALICE:2017iga,ALICE:2019kno} was interpreted as the presence of collective flow in the particle-emitting source. To quantify the effect, the $m_\mathrm{T}$ dependence of the three-dimensional radii~\cite{ALICE:2017iga} was fitted using a power-law function $\sim m_\mathrm{T}^b$~\cite{Kisiel:2014upa} with $b$ characterizing the strength of the flow. The geometrical properties of the source created in collisions were investigated~\cite{ALICE:2015hvw,ALICE:2017iga} via the trend of one- (1D) and three-dimensional (3D) radii as a function of cube root of the average event multiplicity density of charged particles $\langle {\rm d}N_{\rm ch}/{\rm d}\eta \rangle_{|\eta|<0.5}$ (further also referred as multiplicity and denoted as $\langle {\rm d}N_{\rm ch}/{\rm d}\eta \rangle$). As shown in Refs.~\cite{ALICE:2015hvw,ALICE:2017iga}, the radii decrease with decreasing multiplicity manifesting the creation of smaller sources in more peripheral collisions, which is well described by hydrodynamic model scenarios~\cite{Lisa:2005dd}.

In this article, the results of the study of 1D and 3D femtoscopic correlations of two identical charged kaons K$^\pm$K$^\pm$ in Pb--Pb collisions at \mbox{$\sqrt{s_{\rm NN}} = 5.02$}~TeV are presented. This work extends the previous ALICE results in K$^{\pm}$K$^{\pm}$ femtoscopy obtained in Pb--Pb collisions at \mbox{$\sqrt{s_{\mathrm{NN}}} = 2.76$}~TeV~\cite{ALICE:2017iga}. The larger data sample allows performing the analysis in narrower multiplicity ranges providing a more detailed picture of the evolution of the heavy-ion collision. This gives more information on how the source size varies with multiplicity and, as a consequence, allows for tuning and constraining the hydrodynamic models describing the evolution of a heavy-ion collision. The obtained femtoscopic radii are also compared with the predictions of the integrated hydrokinetic model (iHKM)~\cite{PhysRevC.91.014906,PhysRevC.93.024902}, which employs a more advanced description of the chemical freeze-out stage of the emitting-source evolution~\cite{Naboka:2019vko} in comparison to the previous version of the hydrokinetic model (HKM)~\cite{Shapoval:2014wya} used before in Ref.~\cite{ALICE:2017iga}.

One more interesting feature of 3D femtoscopy studies concerns the possibility to investigate the time evolution of the particle emission process as proposed in Refs.~\cite{Shapoval2020,SINYUKOV2016227}. It can be characterized by the time of maximal emission $\tau$, which is related to the instant when particle emission from the source is maximal. It was studied by the ALICE Collaboration for pions and kaons in Pb--Pb collisions at \mbox{$\sqrt{s_{\mathrm{NN}}} = 2.76$}~TeV~\cite{ALICE:2017iga} and compared with the HKM predictions in the 0--5\% centrality range. The comparison showed the importance of taking into account kaon rescattering through the K$^*$(892) resonance. This work extends this study for kaons to several centrality classes to see how the kaon-emission duration changes from central to peripheral collisions. The obtained $\tau$ values are also compared with the iHKM calculations~\cite{Shapoval:2021fqg}.

\section{Detector overview and data selection}
\subsection{ALICE detector, event and track selection}\label{Detector}

The analyzed data sample was recorded by ALICE~\cite{ALICE:2014sbx} in the 2015 LHC Run 2 in Pb--Pb collisions at a center-of-mass energy per nucleon--nucleon pair of \mbox{$\sqrt{s_{\rm NN}} =5.02$}~TeV. Approximately 267 million events were analyzed. They were obtained using the minimum-bias trigger~\cite{ALICE:2018tvk}, which required coincident signals in both V0 scintillators~\cite{ALICE:2015juo,ALICE:2004ftm} to be synchronous with the beam crossing time defined by the LHC clock. Events with multiple primary vertices identified with the silicon pixel detector~\cite{ALICE:1999cls} are tagged as pileup and excluded from the analysis~\cite{ALICE:2014sbx}. The position of the reconstructed event (collision) primary vertex (PV) along the $z$ axis (beam direction) was required to be within 8~cm from the geometrical center of the ALICE detector.

Particles were reconstructed and identified with the central-barrel detectors located within a solenoid magnet with a maximum field strength of ${\rm B}=0.5$~T. The reconstruction of the PV was realized by utilizing the first two inner layers of the Inner Tracking System (ITS)~\cite{ALICE:1999cls,ALICE:2010tia} while tracks and particle momenta were reconstructed in the Time Projection Chamber (TPC)~\cite{ALICE:2000jwd,Alme:2010ke}. The transverse momentum of each track was determined from its bending in the uniform magnetic field. The parameters of the track were determined by performing a Kalman fit to a set of space points (clusters) with an additional constraint for the track to pass through the primary vertex. Only the tracks that have crossed more than 80 out of 159 clusters in the TPC were selected. The fit was required to have a $\chi^2$ per degree-of-freedom lower than 2. Also, for each track, the distance from the reconstructed trajectory to the primary vertex, distance of closest approach (DCA), was required to be less than 0.3~cm in both beam direction and transverse plane in order to reduce the number of secondaries (charged particles produced from the interaction with the detector material, particles from weak decays, etc.). Particle identification (PID) for reconstructed tracks was carried out using the TPC and the Time-Of-Flight (TOF) detectors~\cite{ALICE:2000xcm,ALICE:2002imy,Akindinov:2013tea} in the pseudorapidity range $|\eta| < 0.8$ to avoid regions of the detector with limited acceptance. Charged-kaon tracks were accepted if their transverse momentum was in the range of $0.14<p_{\rm T}<1.50$~GeV$/c$. PID in the TPC was performed by simultaneous measurements of energy loss ($\mathrm{d} E/ \mathrm{d} x$) which is calculated using a parametrized Bethe--Bloch formula, charge, and momentum of each particle passing through the gas chamber. Particle identification in TOF was based on calculation of the time that it takes for a particle with a given mass and momentum to pass through the detector. To be selected, the track was requested to have the measured $\mathrm{d} E/ \mathrm{d} x$ or time of flight within a chosen $N_\sigma$ range for each PID method, where $N_\sigma$ is the number of standard deviations of the measured value from the most probable one~\cite{ALICE:2014sbx,Akindinov:2013tea}. Since the PID performance depends on the momentum, different $N_\sigma$ values were chosen for different track momenta regions.

\subsection{Charged-kaon selection}\label{ChargedKaonSelection}

The PID for charged kaons was carried out by applying the $N_\mathrm{\sigma}$ selection for the tracks in the TPC and TOF detectors. Different $N_\mathrm{\sigma}$ values for different momentum regions were required in order to select as pure a kaon sample as possible, as shown in Tab.~\ref{tab:tab1}. The charged-kaon purity was investigated in two separate particle momentum $p$ intervals as in the analysis of K$^+$K$^-$ correlations in Pb--Pb collisions at \mbox{$\sqrt{s_{\rm NN}}=2.76$}~TeV~\cite{ALICE:2022mxo}. For \mbox{$p < 0.45$}~GeV$/c$, the d$E/$d$x$ distribution in the TPC was used to estimate the contributions of electrons, pions, kaons, and protons to the full measured signal as described in Refs.~\cite{ALICE:2019kno,ALICE:2022mxo}. The purity of kaon candidates with $p\geq0.45$ GeV$/c$ was studied employing the information from both TPC and TOF detectors. The detectors' response was simulated using the Monte Carlo (MC) data from HIJING~\cite{Wang:1991hta} calculations using GEANT3~\cite{Brun:1994aa} to model particle transport through the ALICE detectors. The availability of information from TOF resulted in a final kaon purity being larger than 99\%.
The pair purity was calculated following Refs.~\cite{ALICE:2015hvw,ALICE:2017iga,ALICE:2022mxo} by randomly sampling pairs of $p_{\rm T}$ values according to the experimental $p_{\rm T}$ spectra. For each such pair, the pair purity was obtained as the product of the two single-particle purities and was assigned to the corresponding $k_{\rm T}$. This procedure was repeated until sufficient statistical significance was reached. The resulting pair-purity values are higher than 99\% for pairs of identical charged kaons in all considered $k_{\rm T}$ intervals and centrality classes.

\begin{table}[h]
	\renewcommand{\arraystretch}{1.5}
	\caption{\label{tab:tab1} Charged-kaon selection criteria.}
	\begin{center}
		\begin{tabular}{ l  c  c }
		    \hline \hline
		    & $p_{\rm T}$ & {$0.14 < p_{\rm T} < 1.5$~GeV$/c$} \\
		    & $| \eta |$ & {$< 0.8$} \\
		    & ${\rm DCA}_{\rm xy}$ & {$< 0.3$~cm} \\
		    & ${\rm DCA}_{\rm z}$  & {$< 0.3$~cm} \\\cline{2-3}
           for $p < 0.4$~GeV$/c$: & $|N_{\sigma, \; \mathrm{TPC}}|$ & $< 2$  \\\cline{2-3}
		   for $0.4 \leqslant p < 0.45$~GeV$/c$: & $|N_{\sigma, \; \mathrm{TPC}}|$ &  $< 1$ \\\cline{2-3}
		   \multirow{2}{*}{for $0.45 \leqslant p < 0.5$~GeV$/c$:} & \multicolumn{1}{c}{$|N_{\sigma, \; \mathrm{TPC}}|$} & \multicolumn{1}{c}{$< 2$} \\
		   & \multicolumn{1}{c}{$|N_{\sigma, \; \mathrm{TOF}}|$} & \multicolumn{1}{c}{$< 2$} \\\cline{2-3}
		   \multirow{2}{*}{for $0.5 \leqslant p < 0.8$~GeV$/c$:} & \multicolumn{1}{c}{$|N_{\sigma, \; \mathrm{TPC}}|$} & \multicolumn{1}{c}{$< 3$} \\
		   & \multicolumn{1}{c}{$|N_{\sigma, \; \mathrm{TOF}}|$} & \multicolumn{1}{c}{$< 2$} \\\cline{2-3}
		   \multirow{2}{*}{for $0.8 \leqslant p < 1.0$~GeV$/c$:} & \multicolumn{1}{c}{$|N_{\sigma, \; \mathrm{TPC}}|$} & \multicolumn{1}{c}{$< 3$} \\
		   & \multicolumn{1}{c}{$|N_{\sigma, \; \mathrm{TOF}}|$} & \multicolumn{1}{c}{$< 1.5$} \\\cline{2-3}
		   \multirow{2}{*}{for $p \geqslant 1.0$~GeV$/c$:} & \multicolumn{1}{c}{$|N_{\sigma, \; \mathrm{TPC}}|$} & \multicolumn{1}{c}{$< 3$} \\
		   & \multicolumn{1}{c}{$|N_{\sigma, \; \mathrm{TOF}}|$} & \multicolumn{1}{c}{$< 1$} \\\cline{2-3}
		    & Number of TPC clusters & {$\geqslant80$} \\
        & $\chi^2/{\rm NDF}$ for the track fit & $\leqslant2$\\
	       \hline \hline
\end{tabular}
\end{center}
\end{table}

The femtoscopic correlation function (CF) of two identical particles is known to be sensitive to two-track reconstruction effects such as ``splitting'' and ``merging''. ``Splitting'' is caused by the reconstruction of one track as two, while ``merging'', on the contrary, is a reconstruction of two tracks as one. To suppress merging effects, pairs with the pseudorapidity difference $|\Delta \eta|<0.02$ and the azimuth angle difference between the tracks $|\Delta \varphi^*|<0.04$ were rejected. The angle $\varphi^*$ was corrected for bending inside the magnetic field B at a radial distance from the vertex position inside the TPC of 1.2~m~\cite{Gramling:2012xqa,ALICE:2015tra}. Also, all track
pairs sharing more than 5\% of TPC clusters were rejected to remove the splitting effects.

Events were classified according to the collision centrality determined from the measured charged-particle multiplicity using the energy deposition in the V0 detectors. The present analysis was performed in eight centrality classes: 0--5\%, 5--10\%, 10--20\%, 20--30\%, 30--40\%, 40--50\%, 50--70\%, 70--90\%. The most central collisions are represented by the 0--5\% class. The multiplicities corresponding to these centrality classes are taken from Ref.~\cite{ALICE:2019hno} and presented in Tab.~\ref{tab:tab2}. For the 50--70\% (70--90\%) range, the multiplicity was calculated by averaging the $\langle \mathrm{d} N_\mathrm{ch}/\mathrm{d} \eta \rangle$ values for the 50--60\% and 60--70\% (70--80\% and 80--90\%) intervals.

The following $k_\mathrm{T}$ ranges were considered
\begin{itemize}
	\item four $k_\mathrm{T}$ ranges for the first five centrality classes: $[0.2,0.4]$, $[0.4,0.6]$, $[0.6,0.8]$, and $[0.8,1.2]$~GeV$/c$. The large data sample available allows performing the analysis in four $k_\mathrm{T}$ intervals for every centrality class and, as a consequence, investigating in detail the dependence  of femtoscopic observables with $k_\mathrm{T}$.
	\item only two $k_\mathrm{T}$ ranges for the three most peripheral centrality classes: $[0.2,0.5]$ and $[0.5,1.0]$~GeV$/c$ due to the reduced multiplicity in these centrality classes.
	\item two $k_\mathrm{T}$ ranges for all centrality classes: $[0.2,0.5]$ and $[0.5,1.0]$~GeV$/c$ matching the averaged $\langle k_\mathrm{T} \rangle$ of the Ref.~\cite{ALICE:2017iga} analysis in Pb--Pb at \mbox{$\sqrt{s_{\rm NN}} = 2.76$}~TeV for a direct comparison.
\end{itemize}

\begin{table}[h]
	\renewcommand{\arraystretch}{1.5}
	\caption{\label{tab:tab2} Average event charged-particle multiplicity densities at midrapidity corresponding to the chosen centrality classes~\cite{ALICE:2019hno}.
    }
	\begin{center}
		\begin{tabular}{  c  c  }
			\hline \hline
			Centrality class (\%) & $\langle \mathrm{d} N_\mathrm{ch}/\mathrm{d} \eta \rangle$, $|\eta|<0.5$  \\
			\hline
			0--5 & $1943.0 \pm 56.0$ \\
			5--10 & $1587.0 \pm 47.0$ \\
			10--20 & $1180.0 \pm 31.0$ \\
			20--30 & $786.0 \pm 20.0$ \\
			30--40 & $512.0 \pm 15.0$ \\
			40--50 & $318.0 \pm 12.0$ \\
			50--70 & $139.7 \pm 7.0$ \\
			70--90 & $31.2 \pm 2.7$ \\
			\hline \hline
		\end{tabular}
	\end{center}
\end{table}

\section{Femtoscopic correlation function}

The femtoscopic two-particle correlation function is determined experimentally as the ratio
    $C(\mathbf{q}) = A(\mathbf{q})/B(\mathbf{q})$,
where $\mathbf{q}$ is the pair relative momentum calculated in the coordinate system of choice (see Secs.~\ref{1dCF} and~\ref{3dCF} below). The numerator $A(\mathbf{q})$ is the relative-momentum distribution of pairs of particles from the same event and the denominator $B(\mathbf{q})$ is a reference distribution. To avoid any correlations in the reference distribution $B(\mathbf{q})$, it is constructed by mixing particles from different events containing at least one charged kaon and characterized by a similar multiplicity and $z$ position of the PV (as discussed in Sec.~\ref{Detector}). The similarity of the multiplicity is ensured by dividing each primary centrality range (Tab.~\ref{tab:tab2}) into five equidistant sub-ranges to be used for the sampling of mixed-event particles. In addition, the vertices of the mixed events were constrained to be within 1.6~cm of each other in the $z$ direction. The correlation functions were obtained separately for the two different magnetic field orientations in the experiment, and the difference between them was found to be negligible within uncertainties. Therefore, the data for these two polarities were combined for further analysis.

\begin{figure}
	\center
	\includegraphics[width=0.6\textwidth]{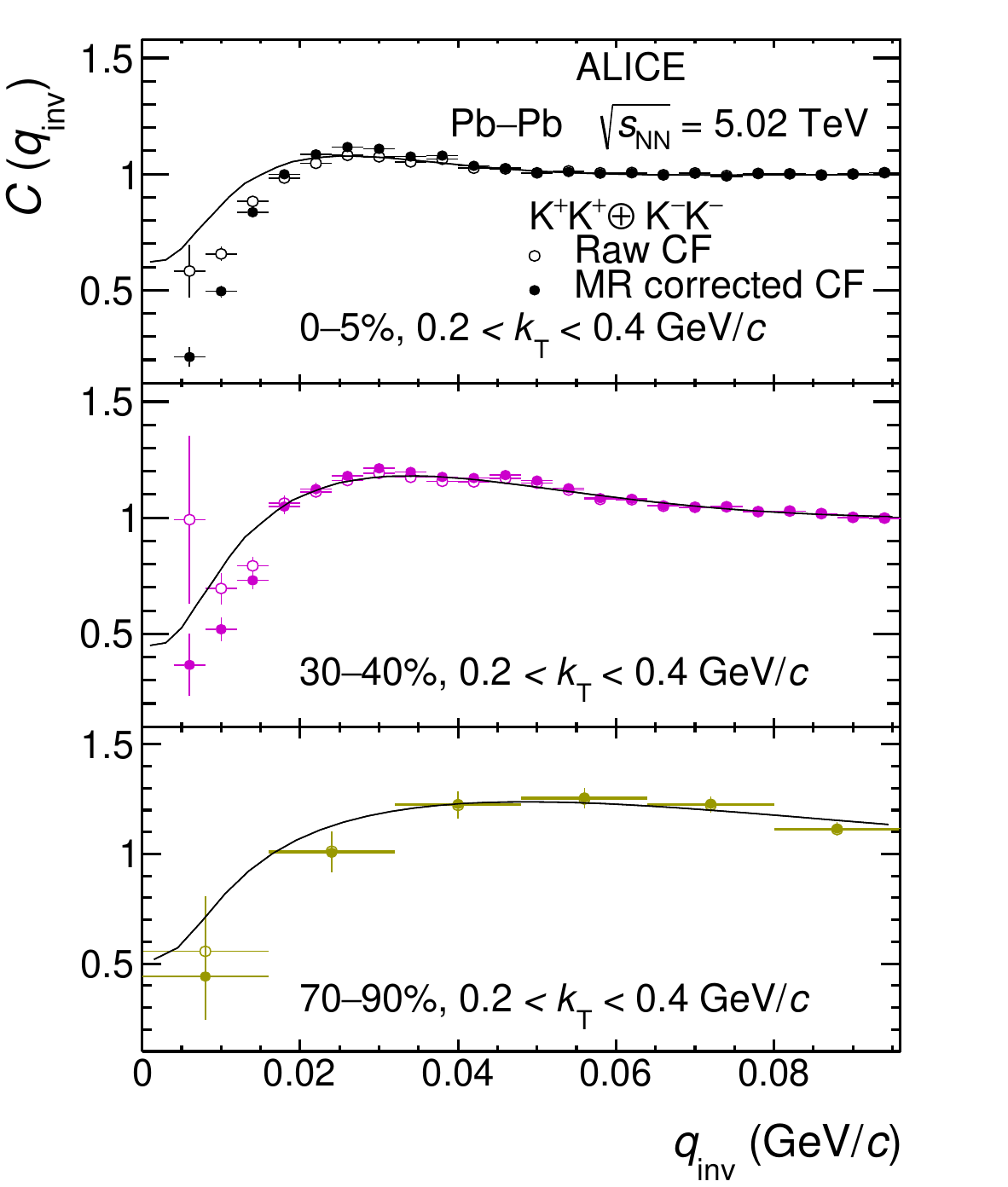}
	\caption{
	1D experimental CFs before (empty markers) and after (solid markers) the momentum resolution (MR) correction given by Eq.~(\ref{MRcorr}) for the 0--5\% (top), 30--40\% (middle), and 70--90\% (bottom) centrality classes. Only CFs in the $k_\mathrm{T}$ [0.2, 0.4]~GeV$/c$ (top and middle panels) and [0.2, 0.5]~GeV$/c$ (bottom panel) ranges are presented.
	\label{fig:DataGraphs/1D-CFs-w-and-wo-res}
	}
\end{figure}

The effect of finite track momentum resolution was accounted by means of the pair relative momentum dependent factor evaluated as a ratio of the correlation functions obtained from MC HIJING simulations taking into account effects of quantum statistics for a given Gaussian source size $R_\mathrm{inv}$ (estimated from a pre-fit of a raw CF) calculated using the ${\rm Lednick\acute{y}}$ code~\cite{Lednicky:2005af} as follows

\begin{itemize}
	\item evaluation of the simulated CF based on particle distributions with the momenta before being propagated through the full simulation of the ALICE detectors from MC HIJING generator for a given Gaussian source;
	\item evaluation of the reconstructed CF based on particle distributions with the momenta after being propagated through the full simulation of the ALICE detectors;
	\item calculation of the correction factor
        \begin{equation}
	       f_{\rm MR} = \frac{C(q_\mathrm{simulated})}{C(q_\mathrm{reconstructed})};\label{MRcorr}
        \end{equation}
	\item correction of the experimental CF by multiplying it by the obtained correction factor $f_{\rm MR}$.
\end{itemize}

The particle distributions for the simulated and reconstructed CFs contain only mixed pairs to exclude two-track effects. In both cases, the correlation weight is calculated using the ${\rm Lednick\acute{y}}$--Lyuboshitz model~\cite{Lednicky:2005af}. The correction factors $f_{\rm MR}$ were calculated as a function of a Gaussian source size with step size of 1~fm and were chosen within $\pm 0.5$~fm range for each experimental CF. The momentum smearing effects are more pronounced in the low relative momentum region, as shown in Fig.~\ref{fig:DataGraphs/1D-CFs-w-and-wo-res}.

\subsection{1D correlation function}\label{1dCF}

In the 1D case, the experimental correlation function $C(q_\mathrm{inv}) := C(\mathbf{q})$ is measured as a function of the invariant two-particle relative momentum

\begin{equation}
    q_\mathrm{inv} = | \mathbf{p_1}^{\rm PRF} - \mathbf{p_2}^{\rm PRF} | ,
\end{equation}

calculated in the Pair Rest Frame (PRF) with $\mathbf{p_1}^{\rm PRF} + \mathbf{p_2}^{\rm PRF} = 0$, where $\mathbf{p_1}^{\rm PRF}$ and $\mathbf{p_2}^{\rm PRF}$ are three-momenta of the particles in a pair.

\begin{figure}[!ht]
	\center{\includegraphics[scale=0.8]{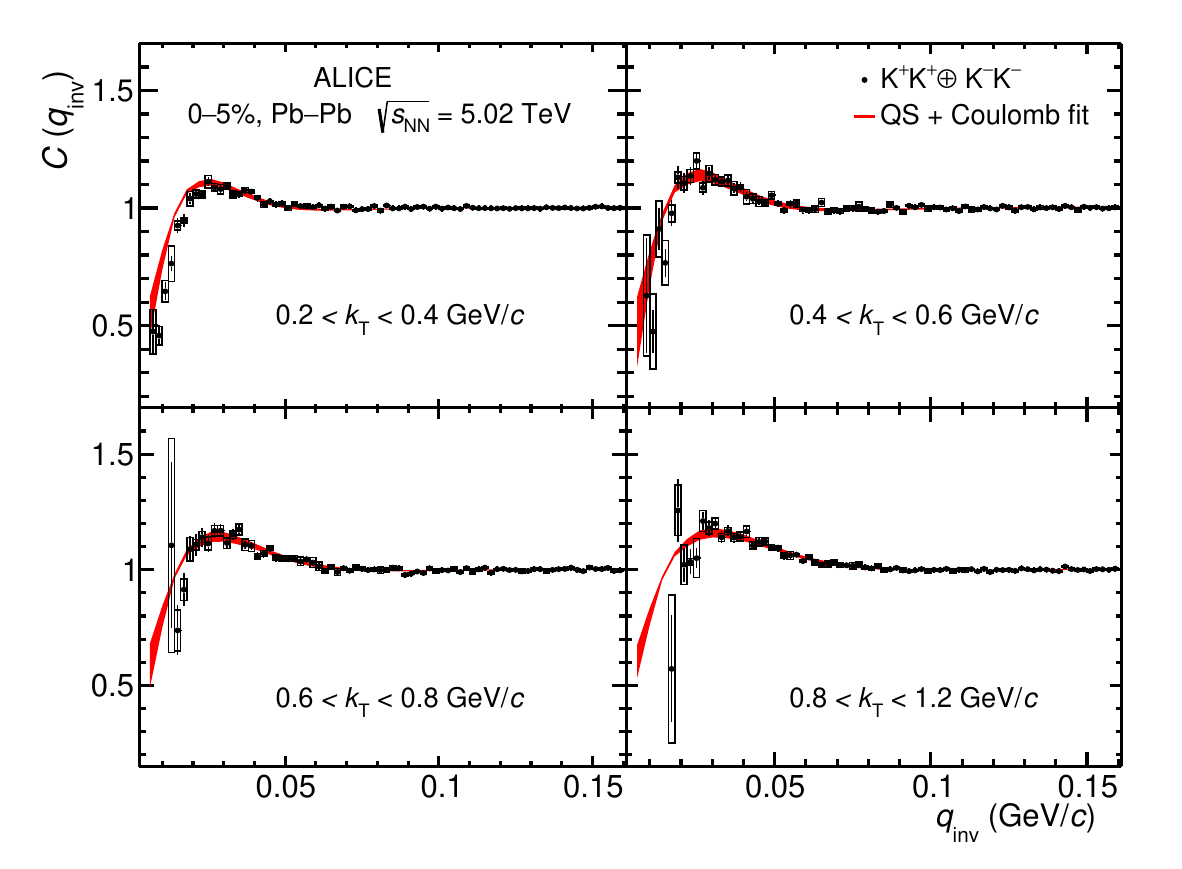}}
	\caption{1D K$^{\pm}$K$^{\pm}$ correlation functions (black markers) for four $k_\mathrm{\rm T}$ ranges and for the 0--5\% centrality class, fitted with Eq.~(\ref{eq1}) (red bands). The CFs are normalized to unity in $0.1<q_\mathrm{inv}<0.15$~GeV$/c$. Statistical uncertainties are shown as bars and systematic uncertainties are shown as boxes. The width of the red bands represents the uncertainties of the Eq.~(\ref{eq1}) fit.}
	\label{figure:DataGraphs/1d_CF_fit_M0_4kt.pdf}
\end{figure}

Assuming a Gaussian-like source, the CF can be parametrized by the Bowler--Sinyukov formula~\cite{BOWLER199169,SINYUKOV1998248}

\begin{equation}
    C(q_\mathrm{inv}) = N \left[ 1 - \lambda +\lambda K \left( R_\mathrm{inv},q_\mathrm{inv} \right) \left( 1 + \exp \left( - R_\mathrm{inv}^2 q_\mathrm{inv}^2 \right) \right) \right],
    \label{eq1}
\end{equation}

where $R_\mathrm{inv}$ is a parameter (called radius) related to a length of homogeneity of the emission source from which the particles with similar momenta are emitted~\cite{AKKELIN1995525,Bowler:1985eny}, $N$ is a normalization factor, $\lambda$ is the correlation strength parameter. The correlation strength is unity in the ideal case of a spherically-symmetric Gaussian source, perfect PID, and the absence of long-lived resonances which decay into kaons. As discussed in Refs.~\cite{ALICE:2014xrc,Akkelin:2001nd}, a suppression of $\lambda$ below unity may also be caused by a finite coherent component of particle emission. The factor $K(R_\mathrm{inv},q_\mathrm{inv})$ describing the Coulomb interaction for a given source radius $R_\mathrm{inv}$ (chosen for each CF from ) is calculated as

\begin{equation}
    K \left( R_\mathrm{inv},q_\mathrm{inv} \right) = \frac{C(\mathrm{QS + Coulomb})}{C(\mathrm{QS})},
    \label{eq2}
\end{equation}

where $C(\mathrm{QS})$ and $C(\mathrm{QS +Coulomb})$ are theoretical correlation functions calculated using the ${\rm Lednick\acute{y}}$--Lyuboshitz model~\cite{Lednicky:2005af} taking into account quantum statistics ``QS'' and quantum statistics and Coulomb interaction ``QS+Coulomb'' effects, respectively.

The parameters $R_\mathrm{inv}$ and $\lambda$ were extracted fitting the data with Eq.~(\ref{eq1}) in the range of the correlation signal $0<q_\mathrm{inv}<0.15$~GeV$/c$. Examples of the one-dimensional experimental correlation functions (for the 0--5\% centrality class) fitted with Eq.~(\ref{eq1}) are shown in Fig.~\ref{figure:DataGraphs/1d_CF_fit_M0_4kt.pdf}. It is also known that CFs for large pair relative momenta can be affected by non-femtoscopic effects such as energy--momentum conservation~\cite{Lisa:2005dd}, resonance decays~\cite{ALICE:2012aai} or mini-jets~\cite{ALICE:2011kmy}, which lead to their deviation from unity. The observed flatness of the CFs obtained in this analysis at $q_{\rm inv}\gtrsim0.08$~GeV$/c$ shows that the non-femtoscopic contributions are negligible.

\subsection{3D correlation function}\label{3dCF}

\begin{figure}
	\center
	\includegraphics[width=0.8\textwidth]{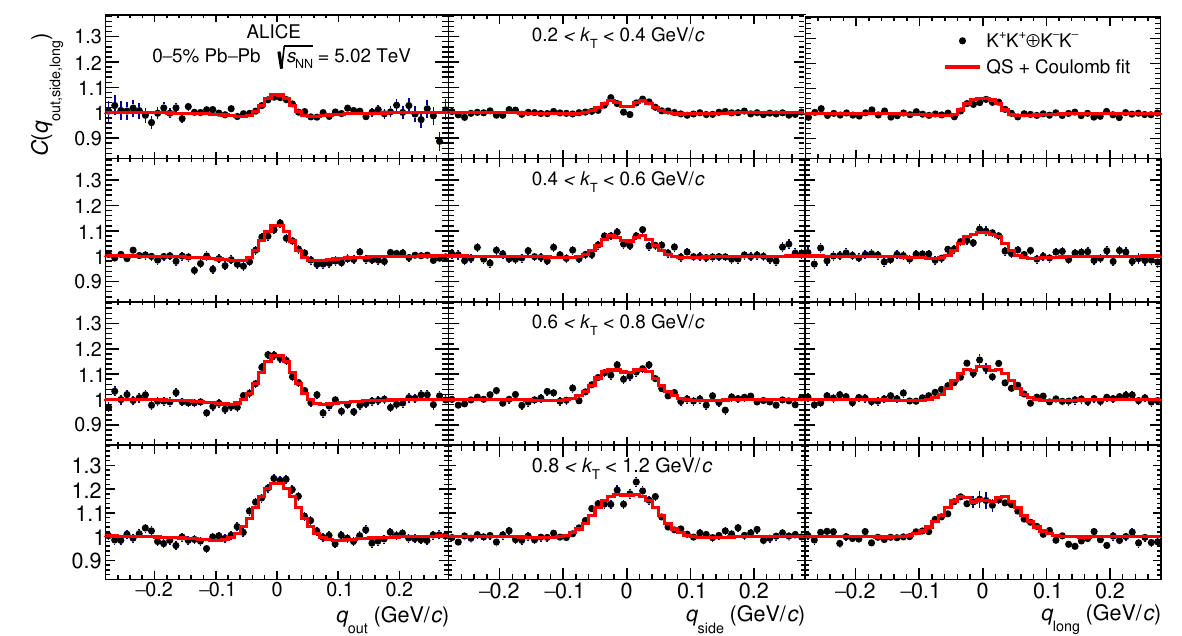}
	\caption{
	3D K$^{\pm}$K$^{\pm}$ correlation functions (black markers) in projections onto the $q_\mathrm{out}$ (left column), $q_\mathrm{side}$ (middle column), and $q_\mathrm{long}$ (right column) axes for four $k_\mathrm{\rm T}$ ranges fitted with Eq.~(\ref{eq3}) (red lines). To project onto $q_\mathrm{i}$ (i$=$out, side, long) component, the others are integrated over the range $|q_\mathrm{i}| < 0.04$~GeV$/c$ to illustrate the femtoscopic effect for the corresponding axis. Statistical uncertainties are shown by bars, and systematic uncertainties are smaller than the markers.
	\label{fig:DataGraphs/3D-0005-CFs-fitted}
	}
\end{figure}

In the 3D case, the correlation function $C(\mathbf{q}_{\rm LCMS}) := C(\mathbf{q})$ is measured as a function of the two-particle relative three-momentum

\begin{equation}
    \mathbf{q}_\mathrm{LCMS} = \mathbf{p_1}^{\rm LCMS} - \mathbf{p_2}^{\rm LCMS},
\end{equation}

where $\mathbf{p_1}^{\rm LCMS}$ and $\mathbf{p_2}^{\rm LCMS}$ are three-momenta of the correlating particles calculated in the Longitudinally Co-Moving System (LCMS) with $p_{1,z}^{\rm LCMS} + p_{2,z}^{\rm LCMS} = 0$~\cite{Lisa:2005dd}. The pair relative momentum $\mathbf{q}_\mathrm{LCMS}$ is decomposed into $q_\mathrm{out}$, $q_\mathrm{side}$, and $q_\mathrm{long}$ components~\cite{Pratt:1986ev,Bertsch:1988db}. The ``long'' component is going along the beam, ``out'' along the pair transverse momentum and ``side'' is perpendicular to the other two.

For the parametrization of the 3D correlation function, the Bowler–Sinyukov formula (for a Gaussian-like source)~\cite{BOWLER199169, SINYUKOV1998248}

\begin{equation}
    C(q) = N \left[ 1 - \lambda + \lambda K \left( R_{\rm inv}, q_{\rm inv} \right) \left( 1 + \exp \left( - R_{\rm out}^2 q_{\rm out}^2 - R_{\rm side}^2 q_{\rm side}^2 - R_{\rm long}^2 q_{\rm long}^2 \right) \right) \right]
    \label{eq3}
\end{equation}

is used, where $R_{\rm out}$, $R_{\rm side}$, and $R_{\rm long}$ are the femtoscopic radii. The $R_{\rm out}$ component is related to the geometrical size of the emission region and particle emission duration, $R_{\rm side}$ to the geometrical size, and $R_{\rm long}$ to the time of maximal emission. The parameters $N$, $\lambda$, and $K(R_{\rm inv}, q_{\rm inv})$ have the same meaning as in Eqs.~(\ref{eq1}) and (\ref{eq2}).

The parameters $R_\mathrm{out,side,long}$ and $\lambda$ were extracted fitting the data with Eq.~(\ref{eq3}) in the range of the correlation signal $-0.15<q_\mathrm{out,side,long}<0.15$~GeV$/c$. As in the 1D case, the non-femtoscopic effects were found to be negligible.
The 3D correlation functions fitted with Eq.~(\ref{eq3}) for the 0--5\% centrality class are shown (in projections) in Fig.~\ref{fig:DataGraphs/3D-0005-CFs-fitted}.

%%%%%%%%%%%%%%%%%%%%%%%%%%%%%%%%

\section{Systematic uncertainties}

Systematic uncertainties on the extracted source parameters were estimated by varying the ranges of kinematic and PID selections as well as by modifications of the fitting procedure. The systematic uncertainty on the event selection was estimated from the variation of the limit on the reconstructed collision vertex $|V_z| < 8$~cm by $\pm12\%$ and also from the effect of the pileup rejection. The former seems to have a noticeable effect on the extracted parameters only in semicentral and peripheral collisions. The latter was investigated by comparing the radii and $\lambda$ parameters when the requirement to reject events with pileup was switched on and off. Removal of the pileup rejection requirement has a negligible effect on the results possibly due to the tight DCA selection criterion applied to the tracks. The uncertainty associated with the single track selection criteria was estimated by using the particle transverse momentum dependent DCA selection
$\mathrm{DCA}_\mathrm{xy} < (0.0105 + 0.0350/p_\mathrm{T}^{1.1})$~cm~\cite{ALICE:2015hav,ALICE:2021szj,ALICE:2022zks} instead of the default $\mathrm{DCA}_\mathrm{xy} < 0.3$~cm one. It was found to affect mostly the extracted $\lambda$ parameters. A similar uncertainty effect on the correlation strength parameter was also observed for the PID selection. The variation of the pair purity values by $\pm5\%$ (with the upper limit equal to 1) influences mainly $\lambda$ and is negligible for the radii. The Gaussian source size used to evaluate the momentum resolution correction factor (Eq.~(\ref{MRcorr})) was varied by $\pm1$~fm to estimate the corresponding systematic uncertainty. There is also an uncertainty related to the CF fitting procedure. It was estimated by varying the radius of the Coulomb interaction (Eq.~(\ref{eq2})) by $\pm 0.5$~fm and the range of the fit by $\pm20\%$. The criteria used for pair selection were varied by up to $\pm20\%$, and their effect on the extracted femtoscopic parameters was found to be negligible.

\begin{table}[ht!]
	
	\renewcommand{\arraystretch}{1.7}
	
	\caption{\label{tab:tab3} Systematic uncertainties for $R_\mathrm{inv, out, side, long}$ and $\lambda_\mathrm{1D, 3D}$ parameters. Separate contributions obtained by variations of track/PID criteria and the fitting procedure together with the total systematic uncertainties obtained from Eq.~(\ref{eq:sys}) are shown.}
	\begin{center}
		\resizebox{1.0\linewidth}{!}{
			\begin{tabular}{ c  c  c  c  c  c  c }
				\hline \hline
				 Sources of systematic uncertainty & $R_\mathrm{inv} (\%)$ & $\lambda_\mathrm{1D} (\%)$ & $R_\mathrm{out} (\%)$ & $R_\mathrm{side} (\%)$ & $R_\mathrm{long} (\%)$ & $\lambda_\mathrm{3D} (\%)$ \\
				\hline
                $V_z$ & 0.3--8.9 & 0.8--31.9 & 1.0--25.9 & 0.4--15.2 & 0.4--12.9 & 0.3--17.2  \\
                Pileup & 0.0--10.5 & 0.5--23.1 & 0.1--4.3 & 0.0--17.8 & 0.0--23.6 & 0.5--17.7 \\
				Single track selection & 0.0--6.9 & 0.2--15.8 & 0.4--23.4 & 0.1--7.3 & 0.0--19.5 & 0.1--12.7 \\
                Purity & 0.0--0.6 & 4.8--10.0 & $\sim 0$ & $\sim 0$ & $\sim 0$ & 5.3--6.1 \\
                Momentum resolution & 0.8--3.0 & 2.0--11.6 & 1.7--5.4 & 2.3--4.9 & 1.7--5.3 & 2.3--10.8 \\
                Coulomb interaction range & 0.0--2.0 & 0.0--5.8 & 0.0--2.4 & 0.0--0.7 & 0.0--1.8 & 0.0--5.6 \\
                Fit range & 0.4--22.5 & 0.8--12.6 & 0.7--35.6 & 0.1--26.7 & 0.7--34.0 & 0.1--11.9 \\
                Total & 2.7--27.0 & 7.8--35.5 & 3.8--45.5 & 3.1--32.6 & 3.4--42.0 & 9.5--26.7 \\
                \hline \hline
		\end{tabular} }
	\end{center}
\end{table}

The Barlow criterion~\cite{Barlow:2002yb} was applied to estimate a statistical significance level of the effect produced by every systematic contribution $i$ as
\begin{eqnarray}
\mathfrak{B}=\frac{|y_0-y_{\rm var}^i|}{\sqrt{\sigma_0^2+{\sigma_{\rm var}^i}^2-2\rho\sigma_0\sigma_{\rm var}^i}}, \label{eq:Barlow}
\end{eqnarray}
where $\sigma_0$ is the statistical uncertainty of the central value $y_0$ extracted from the analysis with the default selection criteria and fit parameters, $\sigma_{\rm var}^i$ is the statistical uncertainty of the value $y_{\rm var}^i$ obtained with some variations $i$ of particle selections and parameters of the fit, $\rho$ characterizes the correlation between $y_0$ and $y_{\rm var}^i$. Combining the individual systematic uncertainties (Tab.\ref{tab:tab3}) in quadrature if $\mathfrak{B}>1$, the total systematic uncertainties on the extracted source parameters were calculated as
\begin{eqnarray}
\Delta_{\rm sys}=\sqrt{\sum_i(y_0-y_{\rm var}^i)^2}. \label{eq:sys}
\end{eqnarray}

All systematic contributions estimated in this analysis are shown in Tab.~\ref{tab:tab3} in terms of the overall spread (minimum--maximum) across all centrality classes and $k_\mathrm{T}$ ranges.

\section{Results and discussion}

\subsection{1D radius and \texorpdfstring{$\lambda$}{lambda} parameters}

\begin{figure}
	\center
	\includegraphics[width=0.4\textwidth]{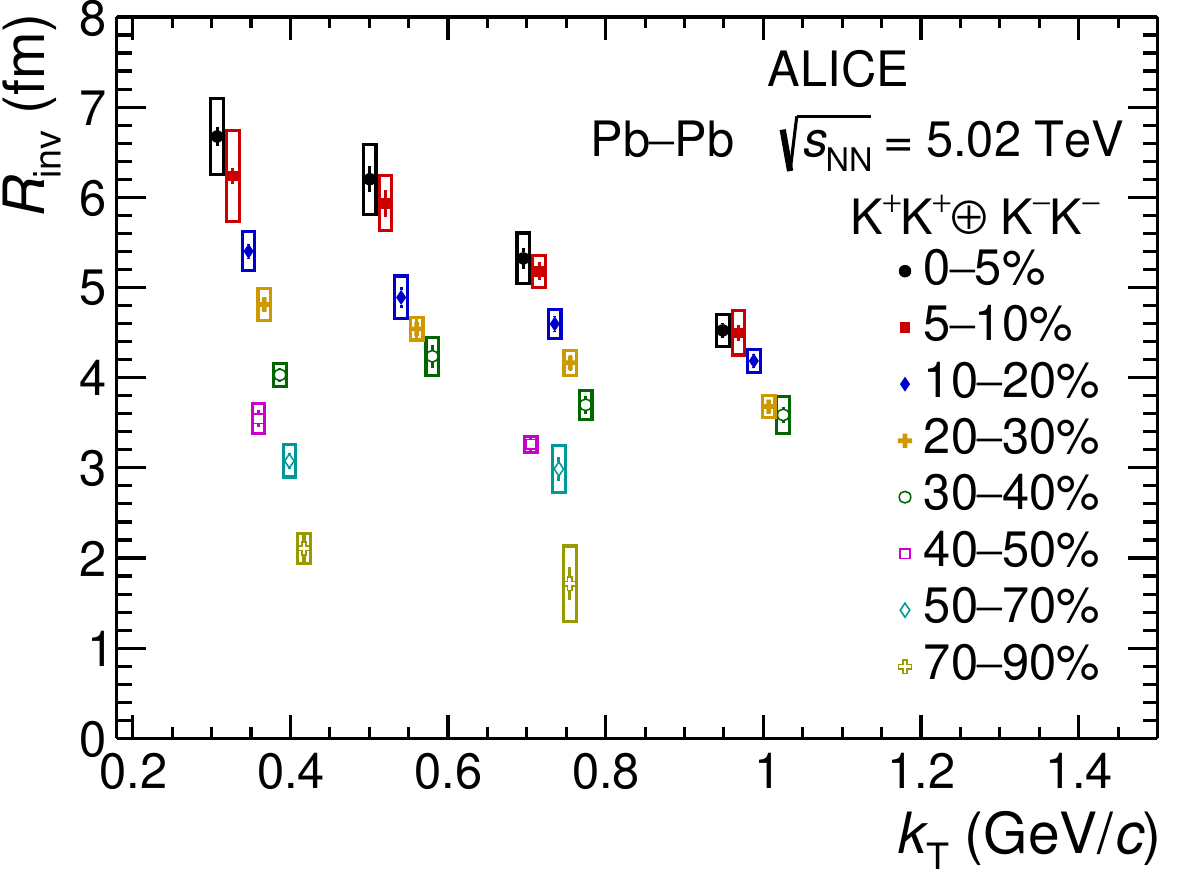}
	 \includegraphics[width=0.4\textwidth]{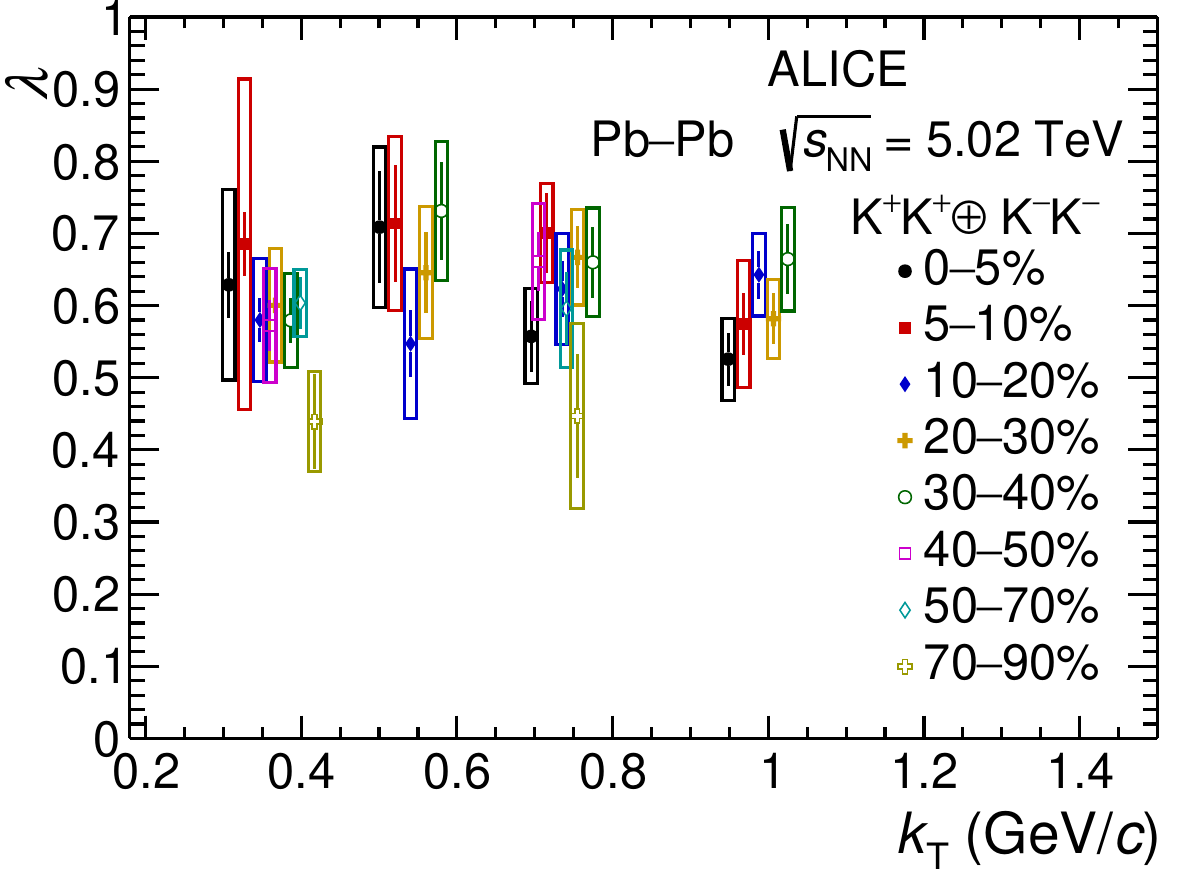}
	\caption{
1D radii (left) and $\lambda$ parameters (right) as a function of the pair transverse momentum $k_\mathrm{T}$. Statistical uncertainties are shown as bars and systematic uncertainties are depicted as boxes. Points for the 5--10\%, 10--20\%, 20--30\%, 30--40\% (50--70\%, 70--90\%) centrality classes are slightly shifted with respect to the 0--5\% (40--50\%) centrality class in the $x$ axis direction for clarity.
	\label{fig:ResGraphs/1D-radii-4kT-all-sys}
	\label{fig:ResGraphs/1D-lambdas-4kT-all-sys}
	}
\end{figure}

Figure~\ref{fig:ResGraphs/1D-radii-4kT-all-sys} (left) shows that the extracted 1D radii $R_\mathrm{inv}$ decrease with increasing $k_\mathrm{T}$. This is a manifestation of the presence of radial flow~\cite{PhysRevC.76.024905} in the system created in Pb--Pb collisions. It can also be noticed that the strength of the flow decreases for more peripheral events since the slope of the $k_\mathrm{T}$ dependence flattens with the decreasing multiplicity. The radii values also decrease with decreasing multiplicity at the same $k_\mathrm{T}$, which indicates that the systems created in more peripheral collisions are smaller.

Figure~\ref{fig:ResGraphs/1D-radii-4kT-all-sys} (right) presents the $k_\mathrm{T}$ dependence of the extracted correlation strength $\lambda$ parameters. A clear suppression below unity is observed for $\lambda$ in all $k_{\rm T}$ intervals and for all centrality classes. As discussed in Sec.~\ref{1dCF}, the possible reasons of this suppression are non-Gaussian features of the correlation function, the influence of K$^*$ and other resonance decays in the final state, and the partially coherent emission of the particles from the source. The figure also demonstrates that the correlation strengths are constant within uncertainties for all multiplicity classes and in all $k_{\rm T}$ ranges.

\begin{figure}
	\center
	 \includegraphics[width=0.4\textwidth]{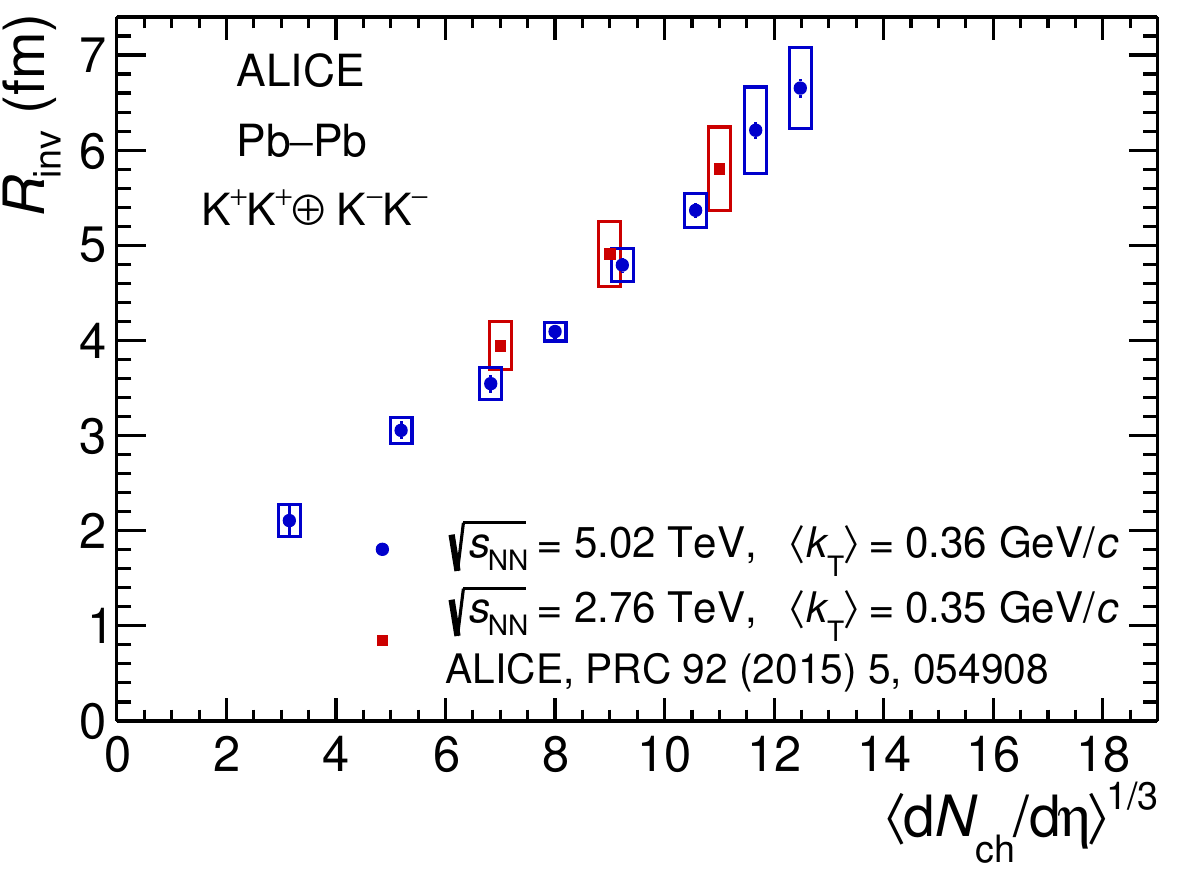}
	 \includegraphics[width=0.4\textwidth]{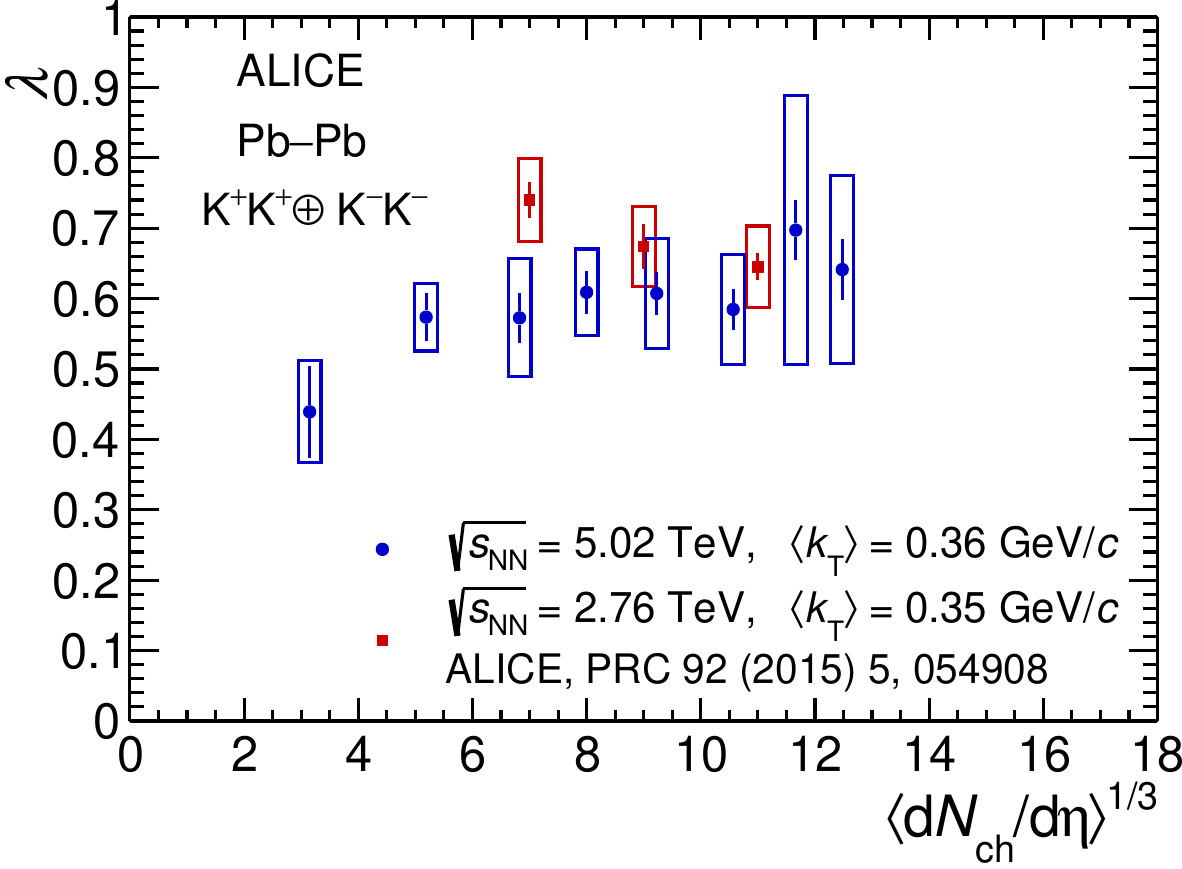}
	\caption{
	1D radii (left) and $\lambda$ parameters (right) as a function of the cube root of the charged-particle multiplicity density $\langle {\rm d}N_{\rm ch}/{\rm d}\eta \rangle^{1/3}$ in the low $k_\mathrm{T}$ range compared with results obtained in Pb--Pb collisions at \mbox{$\sqrt{s_{\rm NN}} = 2.76$}~TeV~\cite{ALICE:2015hvw}. Statistical uncertainties are shown as bars, and systematic uncertainties are shown as boxes.
	\label{fig:ResGraphs/Rinv-276-vs-502-lowkt-all-sys}
	}
\end{figure}

\begin{figure}
	\center
	 \includegraphics[width=0.4\textwidth]{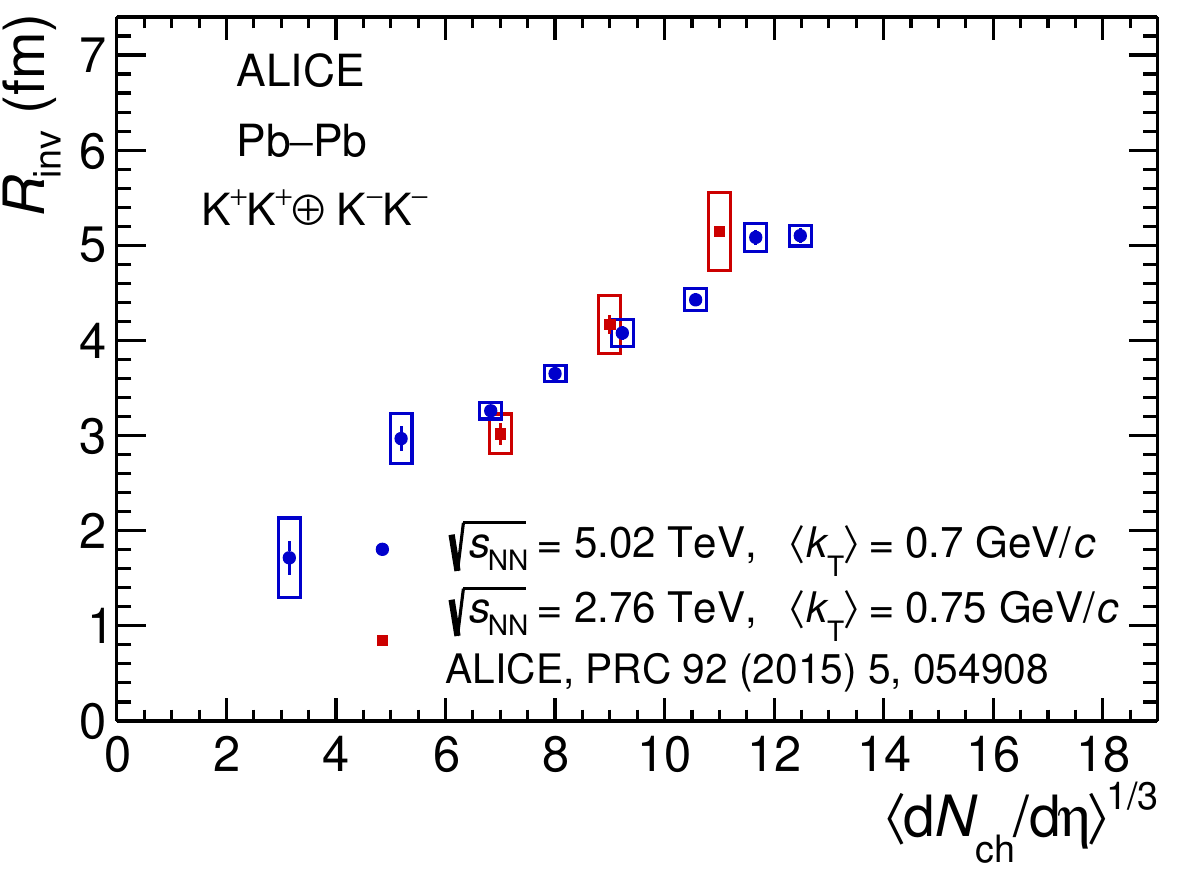}
	 \includegraphics[width=0.4\textwidth]{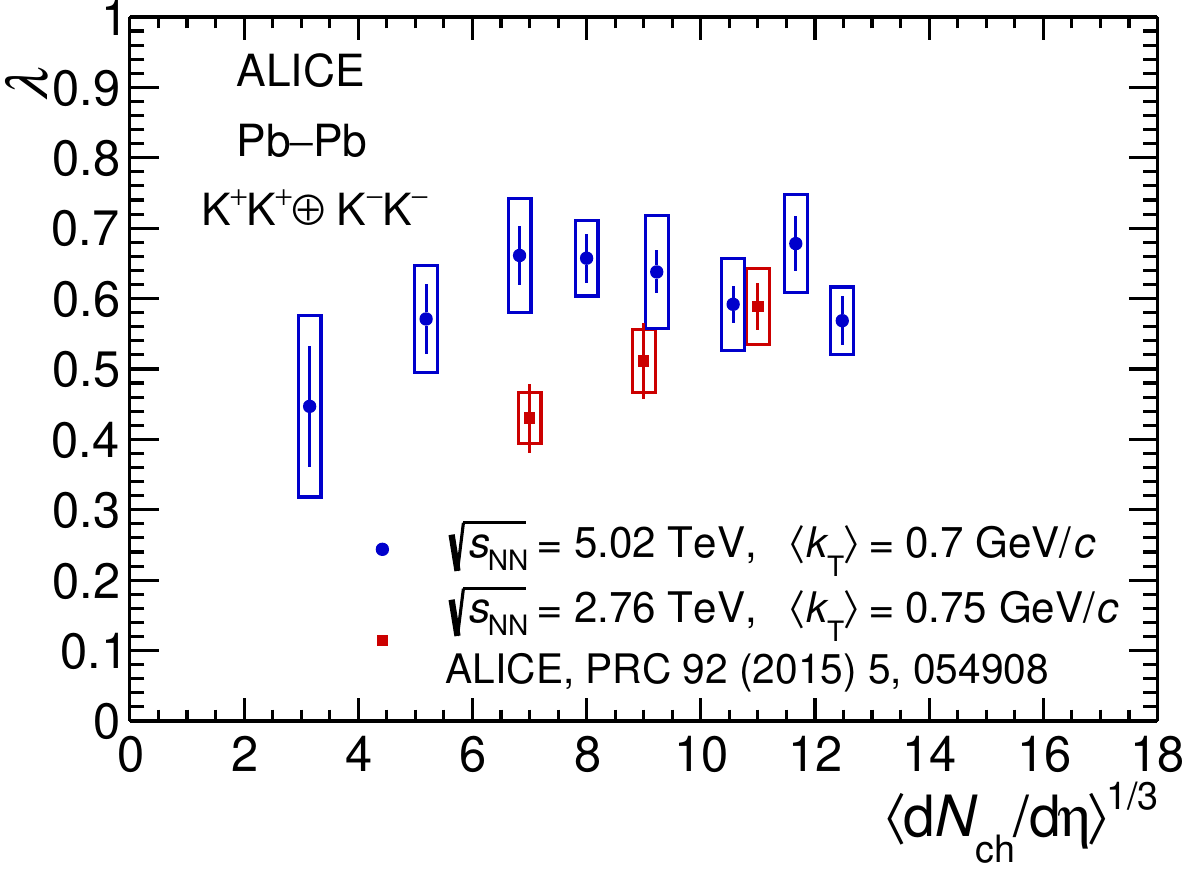}
	\caption{
	1D radii (left) and $\lambda$ parameters (right) as a function of the cube root of the charged-particle multiplicity density $\langle {\rm d}N_{\rm ch}/{\rm d}\eta \rangle^{1/3}$ in the high $k_\mathrm{T}$ range compared with results obtained in Pb--Pb collisions at \mbox{$\sqrt{s_{\rm NN}} = 2.76$}~TeV~\cite{ALICE:2015hvw}. Statistical uncertainties are shown as bars, and systematic uncertainties are shown as boxes.
	\label{fig:ResGraphs/Rinv-276-vs-502-highkt-all-sys}
	}
\end{figure}

In Figs.~\ref{fig:ResGraphs/Rinv-276-vs-502-lowkt-all-sys} and~\ref{fig:ResGraphs/Rinv-276-vs-502-highkt-all-sys}, the K$^{\pm}$K$^{\pm}$ 1D radii extracted in the present analysis are compared with the identical charged-kaon radii obtained in Pb--Pb collisions at \mbox{$\sqrt{s_\mathrm{NN}} = 2.76$}~TeV~\cite{ALICE:2015hvw} as a function of the cube root of the charged-particle multiplicity density. Two pair transverse-momentum ranges with averaged \mbox{${\langle k_\mathrm{T_1}\rangle} = 0.36$}~GeV$/c$ and \mbox{${\langle k_\mathrm{T_2}\rangle} = 0.7$}~GeV$/c$ were considered for the comparison and were chosen to be close to those used in Ref.~\cite{ALICE:2015hvw}. As seen from the comparison, the 1D radii and $\lambda$ parameters obtained in this work agree within uncertainties with the ones from Ref.~\cite{ALICE:2015hvw}.

\subsection{3D radius and \texorpdfstring{$\lambda$}{lambda} parameters}

\begin{figure}[!h] \center
	\begin{tabular}{c}
	   \begin{minipage}[c]{0.45\linewidth}
	       \center{\includegraphics[scale=0.37]{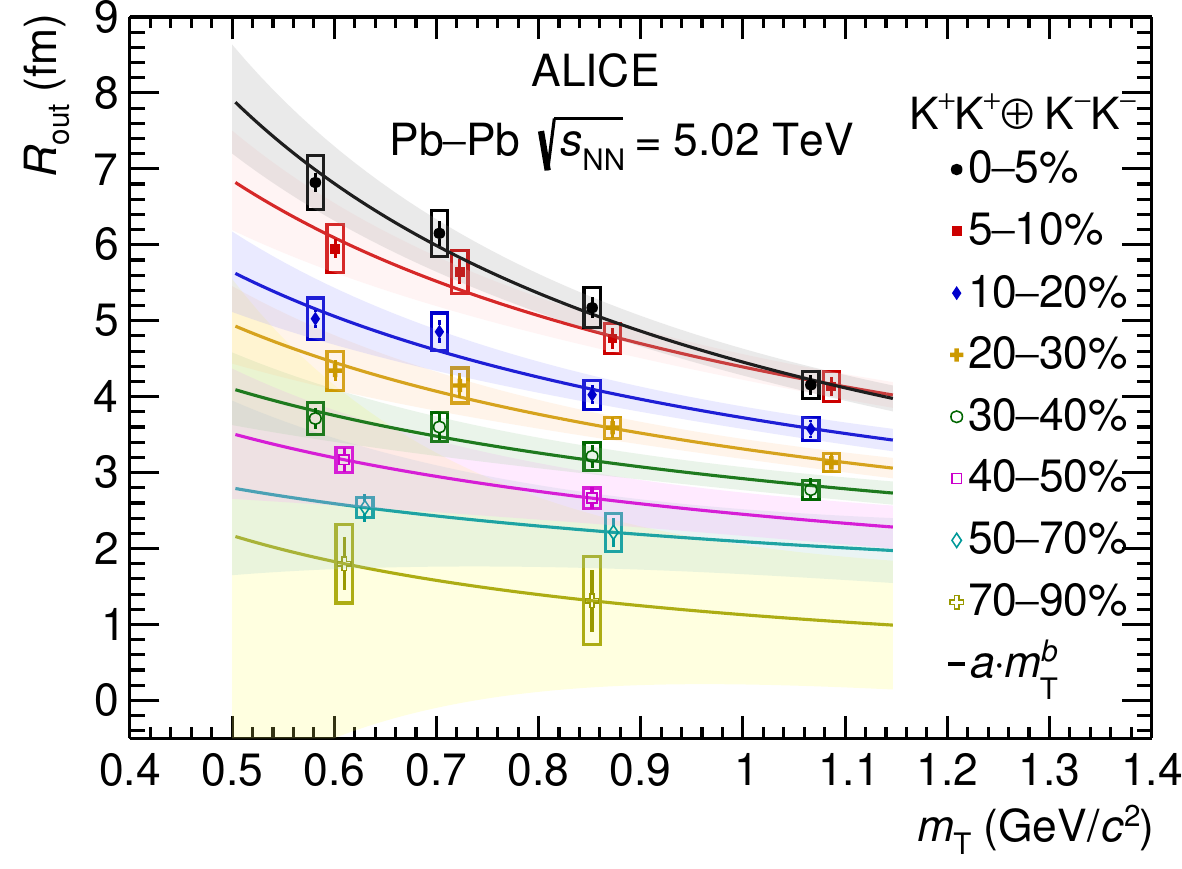}}
	   \end{minipage}
	   \hspace{0.01\linewidth}
	   \begin{minipage}[c]{0.45\linewidth}
	       \center{\includegraphics[scale=0.37]{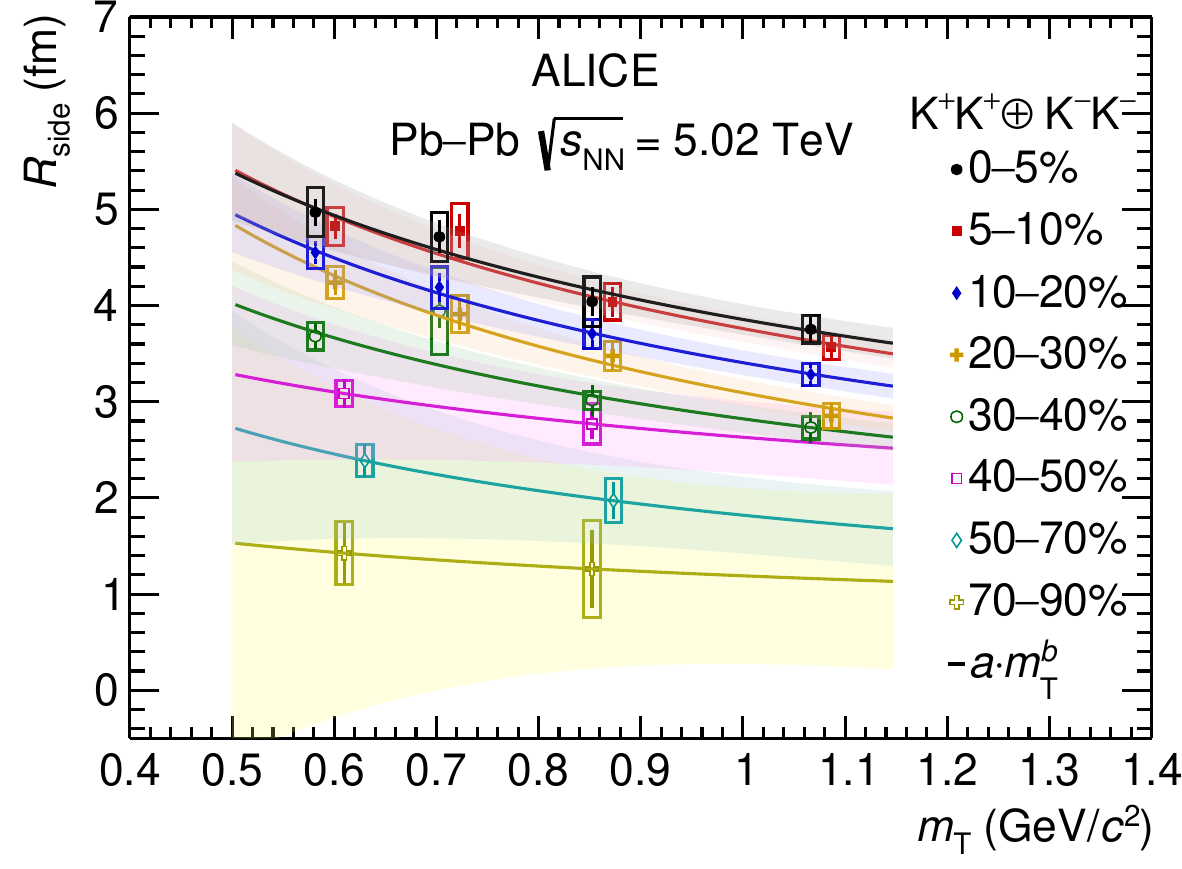}}
	   \end{minipage}
	   \\
	   \begin{minipage}[c]{0.45\linewidth}
	       \center{\includegraphics[scale=0.37]{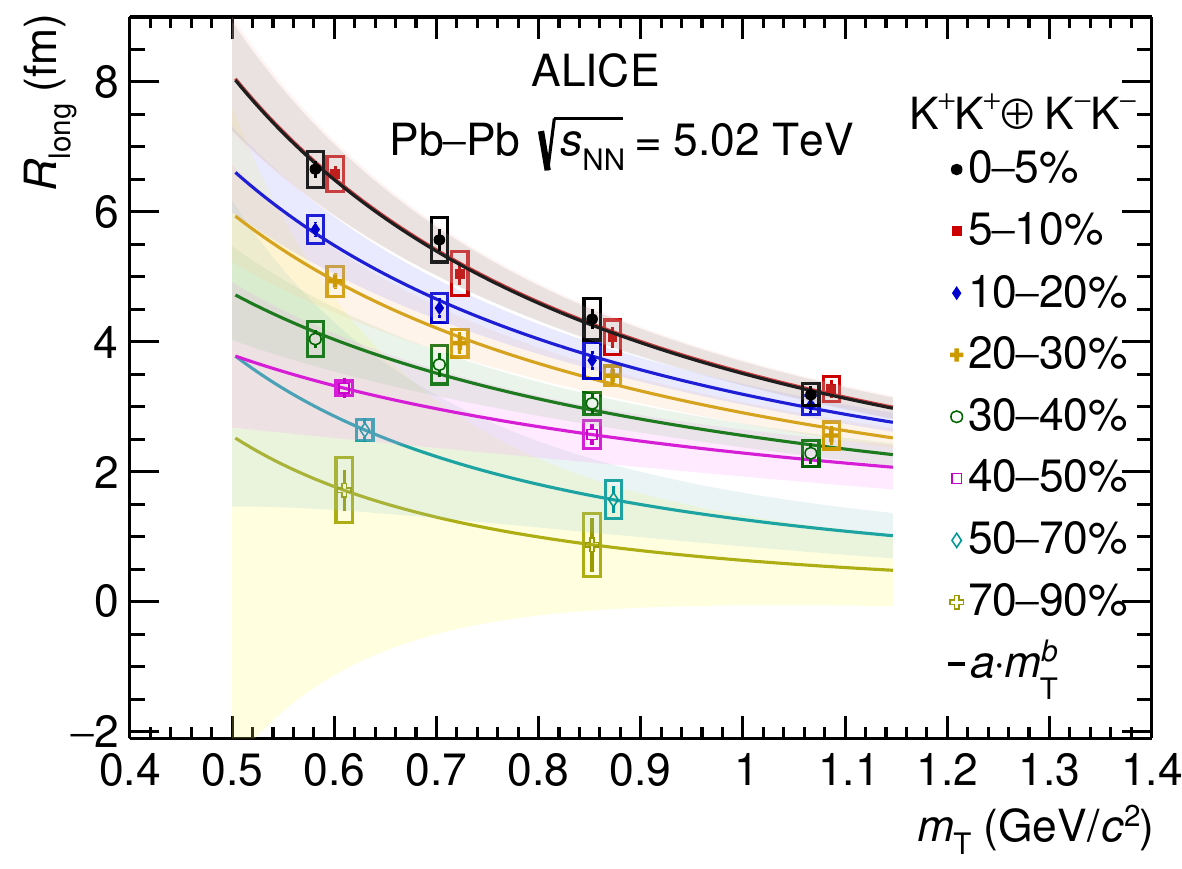}}
	   \end{minipage}
	   \hspace{0.01\linewidth}
	   \begin{minipage}[c]{0.45\linewidth}
	       \center{\includegraphics[scale=0.37]{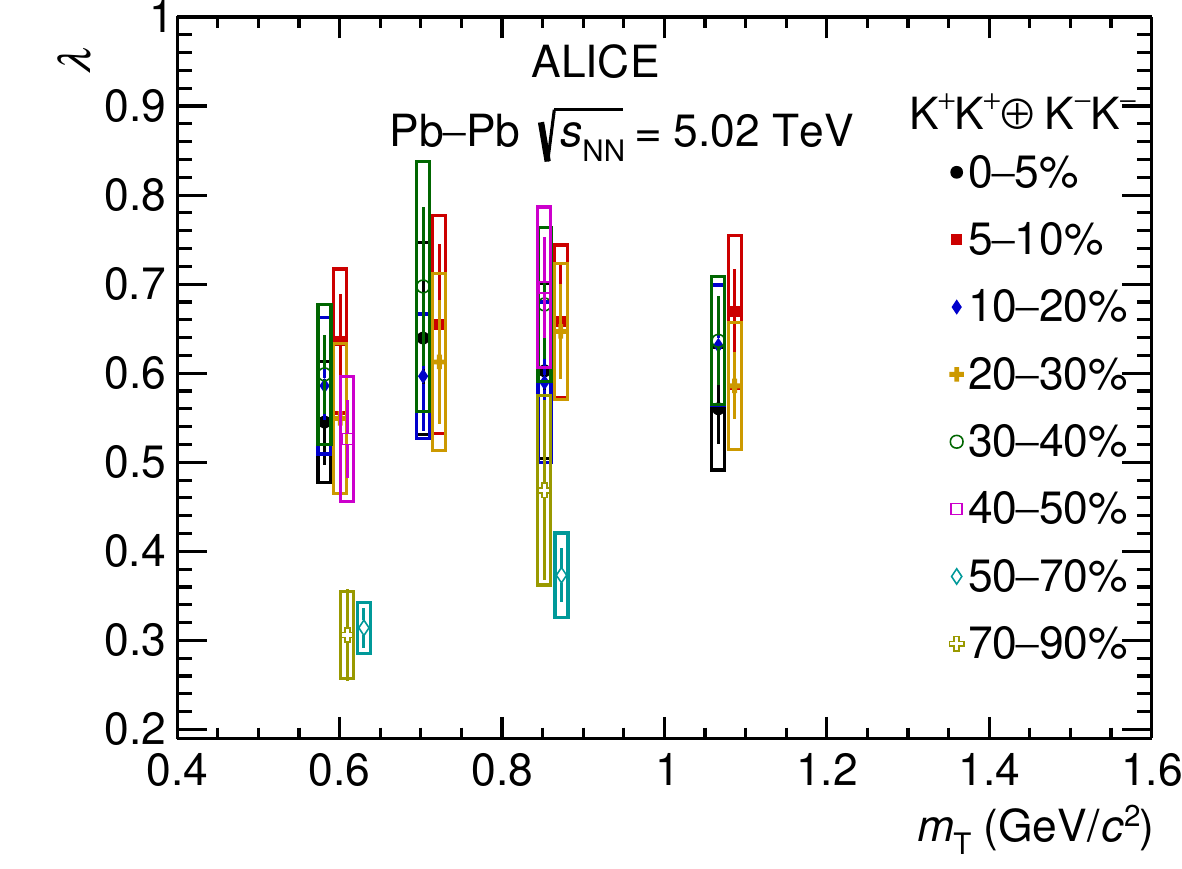}}
	   \end{minipage}
	\end{tabular}
	\caption{K$^{\pm}$K$^{\pm}$ 3D radius and $\lambda$ parameters as a function of the pair transverse mass $m_\mathrm{T}$ fitted with a power-law function $R_\mathrm{out, side, long}(m_\mathrm{T}) = a \cdot m_\mathrm{T}^b$. The data points for the 5--10\%, 20--30\%, and 50--70\% centrality classes are slightly shifted in the $x$ direction for clarity. Statistical uncertainties of the experimental points are shown with bars, and systematic uncertainties are depicted with boxes. The width of the bands represents the uncertainties of the power-law fit.}
	\label{figure:ResGraphs/Rout-mT-fitted-all-sys}
\end{figure}

The 3D radii and $\lambda$ parameter extracted in the present analysis are shown in Fig.~\ref{figure:ResGraphs/Rout-mT-fitted-all-sys} as a function of the pair transverse mass $m_\mathrm{T}$ instead of $k_\mathrm{T}$, which is more convenient for the comparison with the majority of other 3D femtoscopic analyses~\cite{STAR:2001gzb,STAR:2004qya,PHENIX:2009ilf,STAR:2009fks,Kisiel:2014upa} and also necessary for the physics discussion below. As seen in the figure, the radii decrease with increasing $m_\mathrm{T}$. By fitting the $m_\mathrm{T}$ dependence of the radii with a power-law function ${R_\mathrm{out, side, long}(m_\mathrm{T}) = a \cdot m_\mathrm{T}^b}$~\cite{Kisiel:2014upa}, one can extract information about the dynamics of the particle emission process~\cite{Makhlin:1987gm,Wiedemann:1995au}. Table~\ref{tab:tab4} presents the obtained fit parameters $a$ and $b$ for the $R_{\rm out}$, $R_{\rm side}$, and $R_{\rm long}$ projections of the 3D radius. The table shows that the coefficient $a$ decreases for more peripheral events, which is consistent with the decrease of initial source size as predicted, for example, by hydrodynamic scenarios. The power parameter $b$ does not show any dependence on the centrality within the uncertainties of the fit.

\begin{table}[h]
	\renewcommand{\arraystretch}{1.9}
	\caption{\label{tab:tab4} Results of the power-law fit of the 3D radii presented in Fig.~\ref{figure:ResGraphs/Rout-mT-fitted-all-sys}, $\chi^2/\mathrm{NDF} \sim 0$ in the last three centrality ranges because the fit was performed over two data points.}
	\begin{center}
		\resizebox{1.0\linewidth}{!}{
			\begin{tabular}{ c  c  c  c  c  c  c  c  c  c }
				\hline \hline
				Centrality & \multicolumn{3}{ c }{$R_\mathrm{out}$ (fm)} & \multicolumn{3}{ c }{$R_\mathrm{side}$ (fm)} & \multicolumn{3}{ c }{$R_\mathrm{long}$ (fm)} \\
				\cline{2-10}
				(\%)
				& $a$ & $b$ & $\chi^2/\mathrm{NDF}$ & $a$ & $b$ & $\chi^2/\mathrm{NDF}$ & $a$ & $b$ & $\chi^2/\mathrm{NDF}$ \\
				\hline
				0--5 & 4.46 $\pm$ 0.15 & -0.83 $\pm$ 0.010 & 1.05/2 & 3.85 $\pm$ 0.15 & -0.48 $\pm$ 0.12 & 0.47/2 & 3.52 $\pm$ 0.17 & -1.20 $\pm$ 0.11 & 0.63/2 \\
                5--10 & 4.34 $\pm$ 0.14 & -0.62 $\pm$ 0.097 & 1.19/2 & 3.73 $\pm$ 0.14 & -0.52 $\pm$ 0.10 & 1.67/2 & 3.44 $\pm$ 0.16 & -1.18 $\pm$ 0.11 & 0.68/2 \\
                10--20 & 3.72 $\pm$ 0.12 & -0.61 $\pm$ 0.10 & 1.50/2 & 3.41 $\pm$ 0.13 & -0.54 $\pm$ 0.10 & 0.08/2 & 3.18 $\pm$ 0.15 & -1.07 $\pm$ 0.11 & 0.44/2 \\
                20--30 & 3.27 $\pm$ 0.13 & -0.57 $\pm$ 0.13 & 0.48/2 & 3.06 $\pm$ 0.13 & -0.64 $\pm$ 0.12 & 0.79/2 & 2.85 $\pm$ 0.16 & -1.01 $\pm$ 0.14 & 0.83/2 \\
                30--40 & 2.92 $\pm$ 0.14 & -0.49 $\pm$ 0.13 & 0.67/2 & 2.82 $\pm$ 0.14 & -0.51 $\pm$ 0.13 & 1.57/2 & 2.57 $\pm$ 0.16 & -0.88 $\pm$ 0.17 & 0.92/2 \\
                40--50 & 2.45 $\pm$ 0.26 & -0.52 $\pm$ 0.29 & $\sim$ 0 & 2.63 $\pm$ 0.33 & -0.32 $\pm$ 0.31 & $\sim$ 0 & 2.29 $\pm$ 0.35 & -0.73 $\pm$ 0.34 & $\sim$ 0 \\
                50--70 & 2.07 $\pm$ 0.31 & -0.41 $\pm$ 0.39 & $\sim$ 0 & 1.80 $\pm$ 0.29 & -0.57 $\pm$ 0.42 & $\sim$ 0 & 1.23 $\pm$ 0.28 & -1.55 $\pm$ 0.52 & $\sim$ 0 \\
                70--90 & 1.13 $\pm$ 0.81 & -0.95 $\pm$ 1.72 & $\sim$ 0 & 1.19 $\pm$ 0.82 & -0.37 $\pm$ 1.59 & $\sim$ 0 & 0.64 $\pm$ 0.63 & -2.00 $\pm$ 2.17 & $\sim$ 0 \\
                \hline \hline
		\end{tabular} }
	\end{center}
\end{table}

Since $R_\mathrm{side}$ is related to the geometrical size of the system, its behavior can be interpreted similarly to that in the 1D analysis. The behavior of $R_\mathrm{out}$ and $R_\mathrm{long}$ reflects the dynamics of particle emission and evolution duration, respectively. The similarity of the 3D radii dependence on $m_\mathrm{T}$ and the $R_{\rm inv}$ behavior strengthens the conclusion made in the 1D analysis above that systems created in more peripheral collisions are smaller. The extracted values of the 3D $\lambda$ parameter exhibit similar behavior with multiplicity and $m_\mathrm{T}$ to the 1D ones (Fig.~\ref{fig:ResGraphs/1D-radii-4kT-all-sys}). Note that, in the 3D case, a tendency for the $\lambda$ parameter to decrease is observed for peripheral collisions (50--70\% and 70--90\%) compared to central collisions. Such behavior of the parameter $\lambda$ is most likely associated with a larger ratio of $K^*(892)^0/K^\pm$ for non-central events~\cite{ALICE:2021ptz}. This circumstance may result in a greater influence of $K^*(892)^0$ on the extracted source size $K^\pm$, to a larger deviation of its distribution from a Gaussian shape due to the re-scattering processes~\cite{SINYUKOV2016227}. This observation requires further experimental study with a larger data sample.

\begin{figure}
	\center
	 \includegraphics[width=0.4\textwidth]{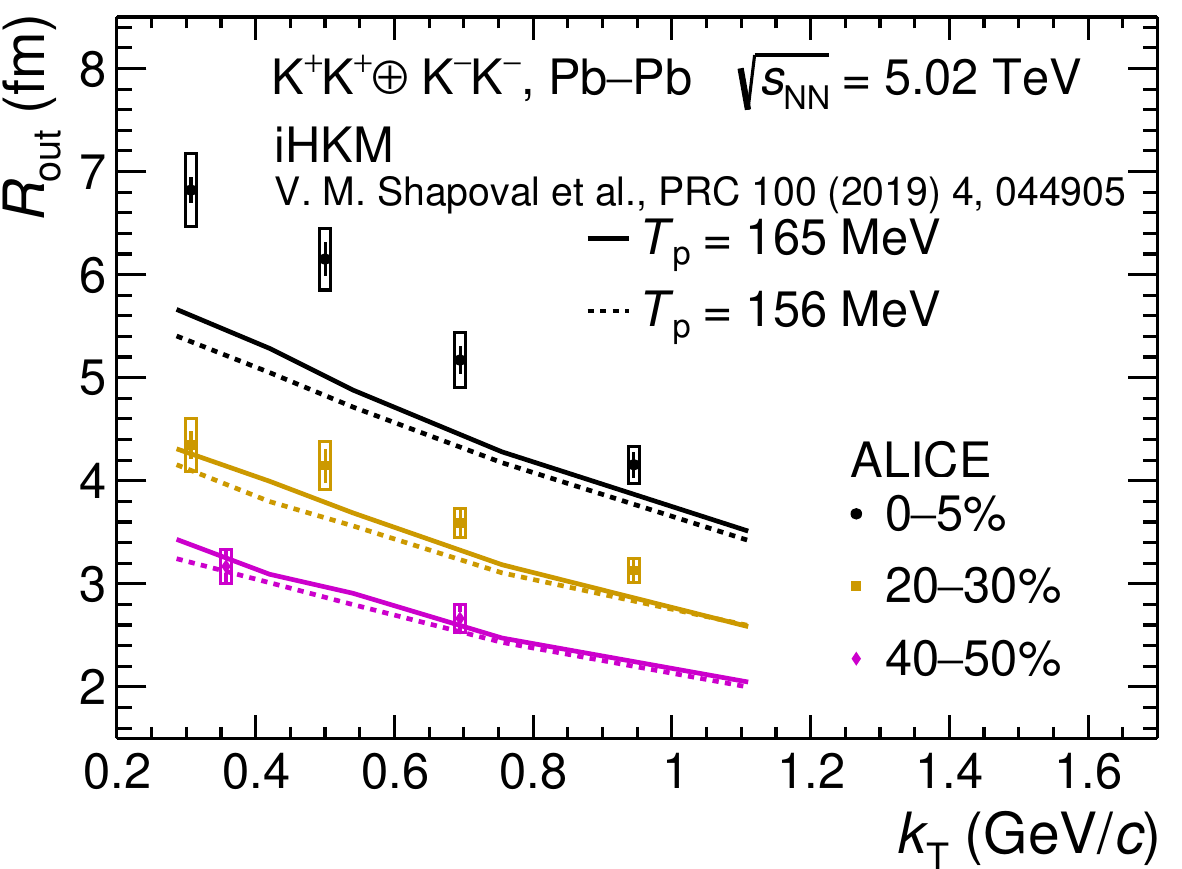}
	 \includegraphics[width=0.4\textwidth]{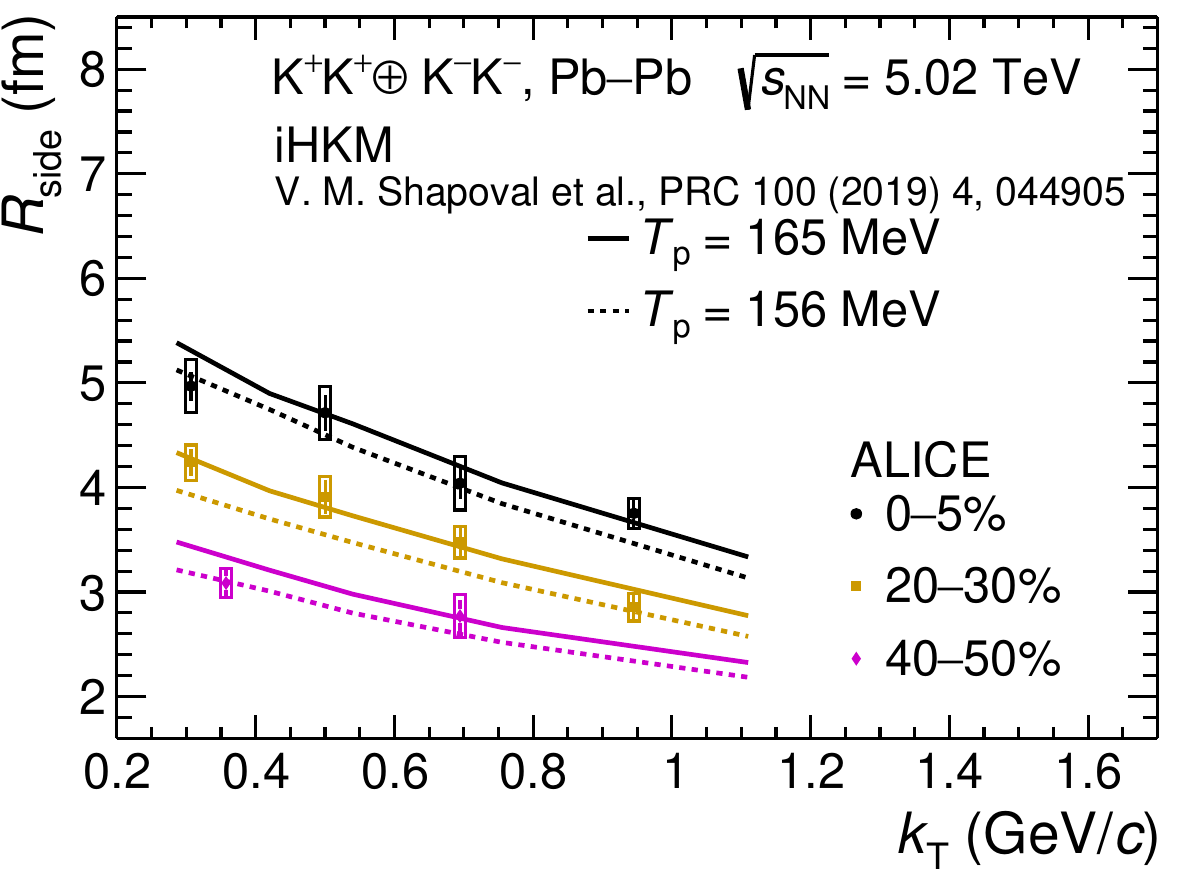}
	 \includegraphics[width=0.4\textwidth]{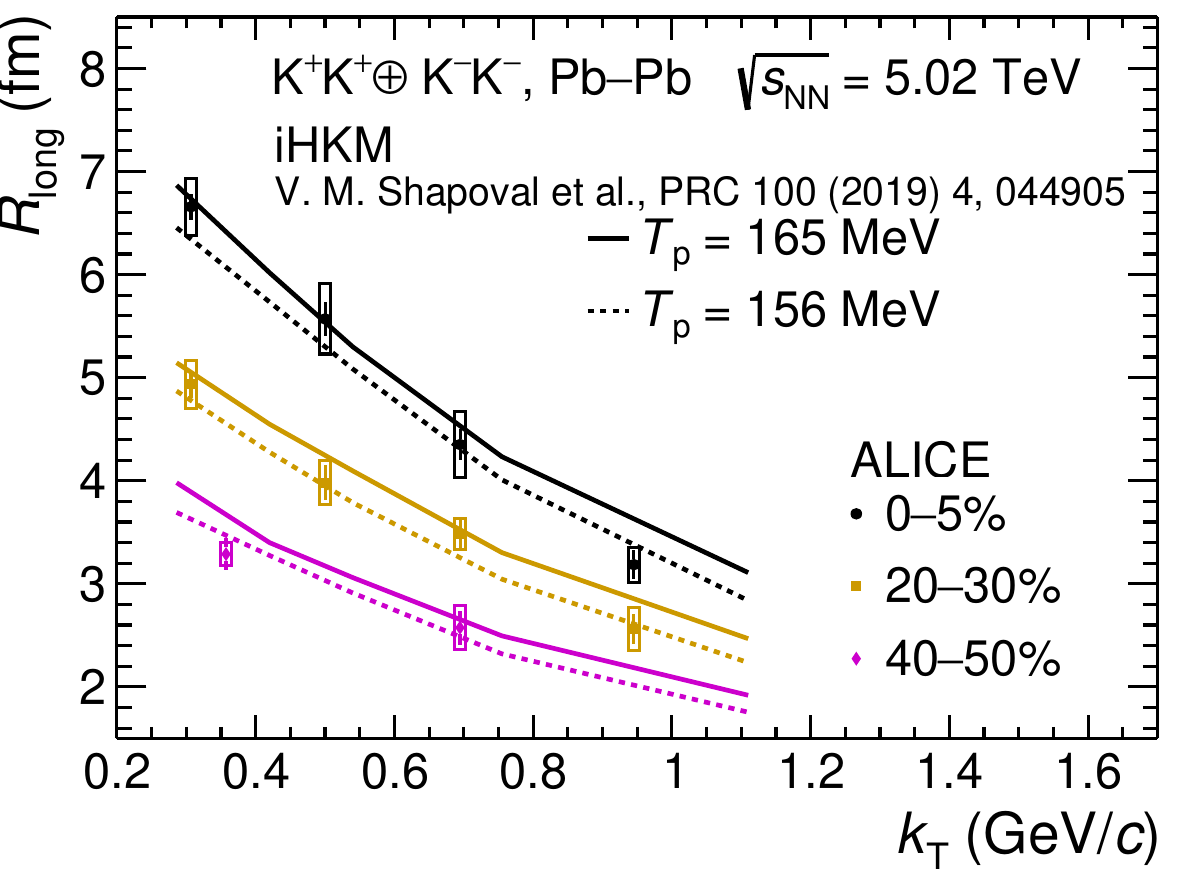}
	\caption{
	Experimental 3D radii (markers) as a function of the pair transverse momentum $k_\mathrm{T}$ compared with iHKM calculations (lines) for the 0--5\%, 20--30\%, and 40--50\% centrality classes. Statistical uncertainties of the experimental points are shown with bars and systematic uncertainties are depicted with boxes. The solid line presents the calculations with the particlization temperature $T_{\rm p}=165$~MeV and the dotted line corresponds to the temperature $T_{\rm p}=156$~MeV.
	\label{fig:ResGraphs/Rout-vs-theory-all-sys}
	}
\end{figure}

The 3D radius parameters extracted in the present analysis are compared with the theoretical predictions of the integrated hydrokinetic model iHKM~\cite{PhysRevC.100.044905,Shapoval2020} for two different equations of state with the corresponding values of the particlization temperature $T_\mathrm{p} = 156$~MeV and $T_\mathrm{p} = 165$~MeV. In the iHKM, the particlization is one of the evolution stages of the hadronic matter created in the collisions, when the QGP transforms into the hadron cascade. The temperature of this transition depends on the utilized equation of state for the QGP (see details in Ref.~\cite{Shapoval2020}). The compared radii are shown as a function of $k_\mathrm{T}$ for three centrality classes: 0--5\%, 20--30\%, and 40--50\% in Fig.~\ref{fig:ResGraphs/Rout-vs-theory-all-sys}. As seen from the figure, the model underestimates the experimental $R_\mathrm{out}$ for the 0--5\% centrality class by about 1.0--1.5~fm. A similar behavior was observed before for $\pi$K correlations in Pb--Pb collisions at \mbox{$\sqrt{s_{\rm NN}}=2.76$}~TeV~\cite{ALICE:2020mkb} in comparison to another hydrodynamic model THERMINATOR~2~\cite{Kisiel:2018wie}. Thus, this remains an open question of femtoscopic studies and probably must be further investigated from a theoretical perspective. Nevertheless, the present data (Fig.~\ref{fig:ResGraphs/Rout-vs-theory-all-sys}) show an overall good agreement with the iHKM predictions, and within the uncertainties, it is not possible to discriminate between the two model parametrizations.

\subsection{Time of maximal emission}

Based on the combined fitting of the $m_\mathrm{T}$ dependence of $R_\mathrm{long}^2$ and the transverse momentum spectrum, one can estimate the time of maximal emission for particles of a given species~\cite{Shapoval2020, SINYUKOV2016227}. In the present work, this method was applied for the extraction of maximal emission time $\tau_{\rm K}$ for kaons.

\begin{figure}
	\center
	\includegraphics[width=0.6\textwidth]{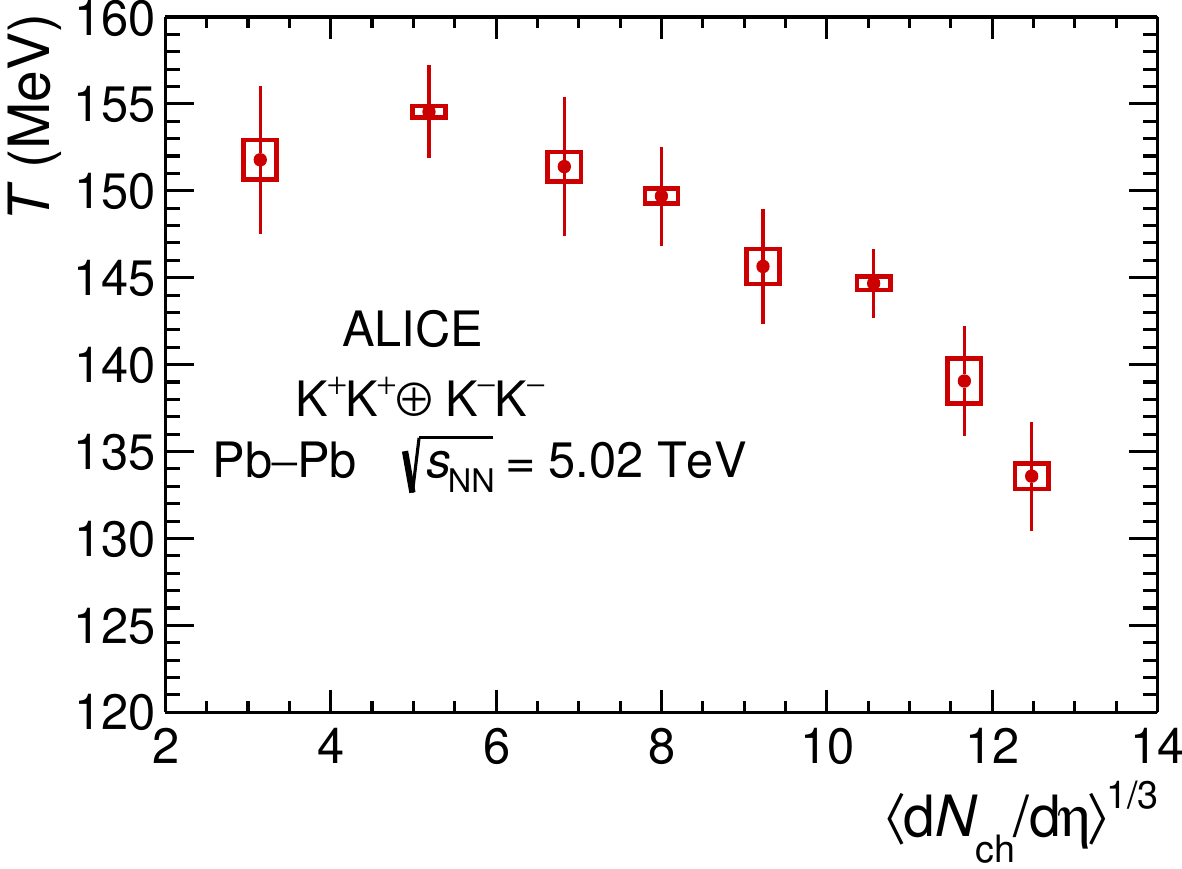}
	\caption{
	Maximal emission temperature $T$ for charged pions and kaons in Pb--Pb collisions at \mbox{$\sqrt{s_{\rm NN}}=5.02$}~TeV as a function of the cube root of the charged-particle multiplicity density \mbox{$\langle {\rm d}N_{\rm ch}/{\rm d}\eta\rangle^{1/3}$}.
Statistical uncertainties are shown as bars and systematic uncertainties are depicted as boxes.
	\label{fig:ResGraphs/T_all_mult}
	}
\end{figure}

The first step is to perform a combined fit of pion and kaon $p_\mathrm{T}$ spectra with the following formula~\cite{Shapoval2020, SINYUKOV2016227}:
\begin{equation}
	E \frac{d^3 N}{d^3 p} \; \propto \; \exp \left[- \left(\frac{m_\mathrm{T}}{T} + \alpha \right) \sqrt{1-\bar{v}^2_{\rm T}} \right],
	\label{eq9}
\end{equation}
where $T$ is the maximal emission temperature, $\bar{v}_\mathrm{T} = \frac{k_\mathrm{T}}{m_\mathrm{T} + \alpha T}$ is the transverse collective velocity, $\alpha$ is a parameter characterizing the strength of collective flow (infinite value means zero flow and small value means strong flow). From the fit one extracts the common effective temperature of pions and kaons $T$ and the two values of $\alpha$ for each particle species. The data on the pion and kaon spectra in Pb--Pb collisions at \mbox{$\sqrt{s_{\rm NN}}=5.02$}~TeV are taken from Ref.~\cite{ALICE:2019hno} and their combined fit was performed in the pion and kaon momentum $p$ ranges of $0.5 < p_\mathrm{\pi, K} < 1.0$~GeV$/c$ as suggested in Refs.~\cite{Shapoval2020, SINYUKOV2016227}. The maximal emission temperatures extracted from the fit are presented in Fig.~\ref{fig:ResGraphs/T_all_mult}. The systematic uncertainties for the obtained values of $T$ were estimated by the variation of the fit range of the spectra by $\pm 20\%$.

\begin{figure}
	\center
	\includegraphics[width=0.6\textwidth]{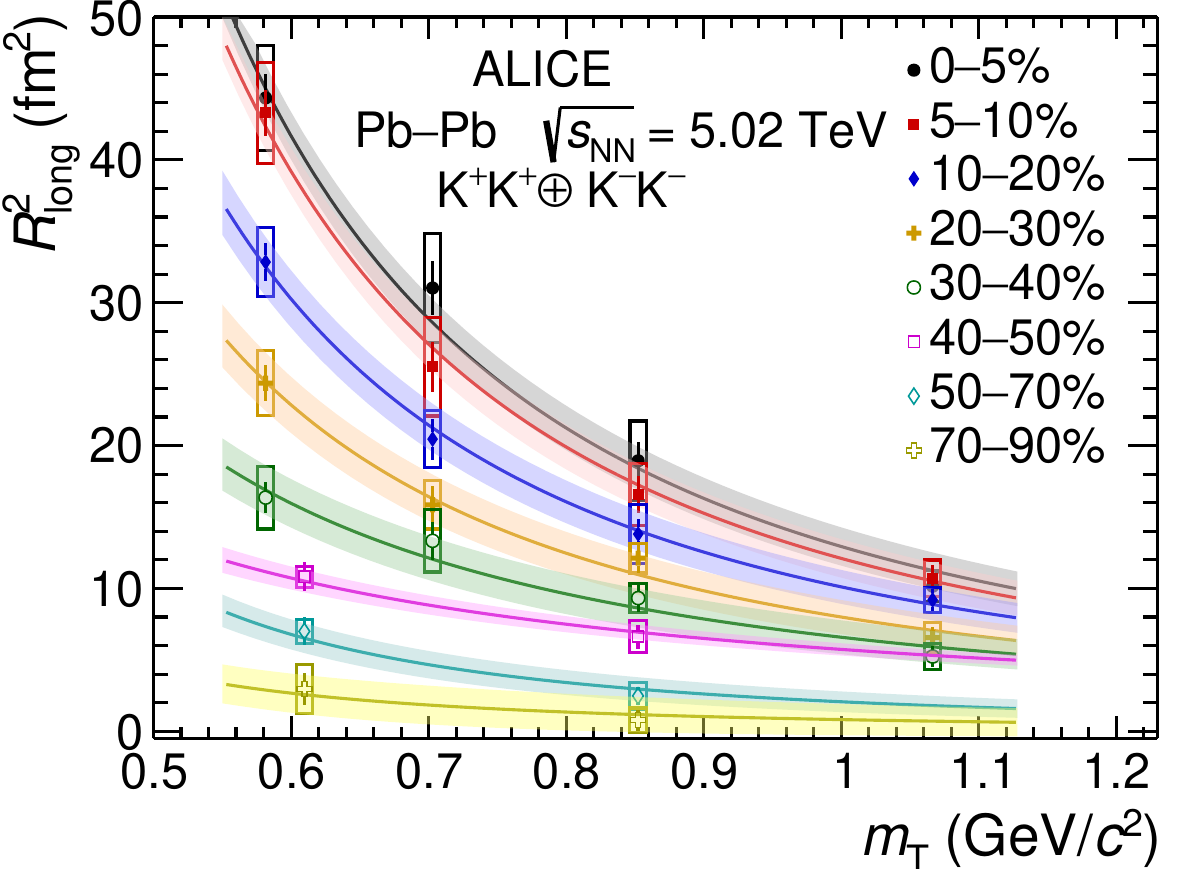}
	\caption{
	$R_\mathrm{long}^2$ fitted with Eq.~(\ref{eq8}) for eight centrality classes: 0--5\%, 5--10\%, 10--20\%, 20--30\%, 30--40\%, 40--50\%, 50--70\%, and 70--90\%. Statistical uncertainties are shown as bars and systematic uncertainties are depicted as boxes. The width of the bands represents the uncertainties of the power-law fit.
	\label{fig:ResGraphs/Rlong2-fit-all-sys}
	}
\end{figure}

The second step is to parametrize the $m_\mathrm{T}$ dependence of $R_\mathrm{long}^2$ using the formula

\begin{equation}
	\begin{split}
		& R_\mathrm{long}^2(m_\mathrm{T})=\tau_{\rm K}^2\zeta^2 \left(1+\frac{3}{2}\zeta^2 \right), \\
		& \zeta^2 = {(\zeta_l/\tau_{\rm K})}^2 = \frac{T}{m_\mathrm{T}}\sqrt{1-\bar{v}^2_{\rm T}},
	\end{split}
	\label{eq8}
\end{equation}

where $\zeta_l$ is the longitudinal homogeneity length, $T$ is taken from the fit of the pion and kaon spectra with Eq.~(\ref{eq9}), and $\alpha$ is a free parameter~\cite{Shapoval2020, SINYUKOV2016227, Shapoval:2021fqg}. The results of the fits to $R_\mathrm{long}^2$ with Eq.~(\ref{eq8}) are shown in Fig.~\ref{fig:ResGraphs/Rlong2-fit-all-sys}.

\begin{figure}
	\center
	\includegraphics[width=0.7\textwidth]{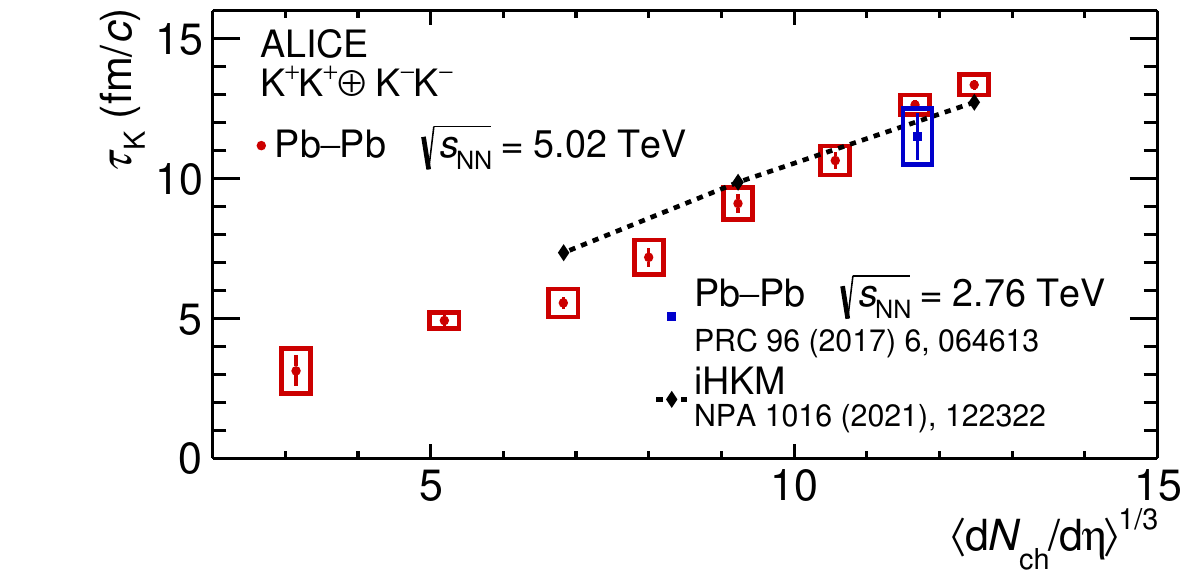}
	\caption{
	Time of maximal emission $\tau_{\rm K}$ for charged kaons in Pb--Pb collisions at \mbox{$\sqrt{s_{\rm NN}}=5.02$}~TeV compared with the iHKM calculations~\cite{Shapoval:2021fqg} and the result of the K$^\pm$K$^\pm$ analysis in Pb--Pb collisions at \mbox{$\sqrt{s_{\rm NN}}=2.76$}~TeV~\cite{ALICE:2017iga}. Statistical uncertainties are shown as bars and systematic uncertainties are depicted as boxes.
	\label{fig:ResGraphs/tau-vs-mult-all-sys}
	}
\end{figure}

The $\tau_{\rm K}$ parameters obtained from the fit with Eq.~(\ref{eq8}) are presented as a function of the cube root of the charged-particle multiplicity density in Fig.~\ref{fig:ResGraphs/tau-vs-mult-all-sys}. Systematic uncertainties of the extracted $\tau_{\rm K}$ were estimated from variation of the temperature $T$ in Eq.~(\ref{eq8}) corresponding in turn to the variation of the spectra fit range. The obtained $\tau_{\rm K}$ is also compared with the iHKM calculations~\cite{Shapoval:2021fqg} for the 0--5\%, 20--30\%, and 40--50\% centrality classes as well as with $\tau_{\rm K}$ extracted in the K$^{\pm}$K$^{\pm}$ femtoscopic analysis in Pb--Pb collisions at \mbox{$\sqrt{s_{\mathrm{NN}}} = 2.76$}~TeV for the 0--5\% centrality class~\cite{ALICE:2017iga}. As can be seen from Fig.~\ref{fig:ResGraphs/tau-vs-mult-all-sys}, the results provided by iHKM agree with the experimental data. The values of maximal emission time $\tau_{\rm K}$ obtained for both the \mbox{$\sqrt{s_{\mathrm{NN}}} = 2.76$}~TeV and \mbox{$\sqrt{s_{\mathrm{NN}}} = 5.02$}~TeV Pb--Pb collisions are also in good agreement within uncertainties.

The maximal emission of particles from the source occurs after the chemical freeze-out, so the obtained values for the emission temperature $T$ are less than the temperature of this freeze-out stage $T_\mathrm{ch} \sim 160$~MeV~\cite{Shapoval2020, SINYUKOV2016227}. Note that a similar combined fit of pion, kaon, and proton $p_\mathrm{T}$ spectra in Pb--Pb collisions at $\sqrt{s_{\rm NN}}=5.02$~TeV performed using the usual blast-wave model gives comparable values for the kinetic freeze-out temperature $T_\mathrm{kin}$, which has slightly stronger centrality dependence~\cite{ALICE:2019hno}. Both $T$ and $T_\mathrm{kin}$ decrease (within uncertainties) with increasing centrality. This can be interpreted as a longer (shorter) expansion of the hadronic matter created in central (peripheral) collisions. The same conclusion can be made from the behavior of the extracted time of maximal emission $\tau_{\rm K}$, which decreases with the decreasing collision multiplicity as seen in Fig.~\ref{fig:ResGraphs/tau-vs-mult-all-sys}.

\section{Summary}

Femtoscopic correlations have been studied with K$^{\pm}$K$^{\pm}$ pairs in Pb--Pb collisions at \mbox{$\sqrt{s_{\rm NN}}=5.02$}~TeV by ALICE. This work significantly complements and extends the results obtained in the previous K$^{\pm}$K$^{\pm}$ femtoscopic analysis in Pb--Pb collisions at \mbox{$\sqrt{s_{\mathrm{NN}}} = 2.76$}~TeV. The larger LHC Run~2 experimental data sample in comparison to that of the Run~1 allowed for a more detailed investigation of both the one- and the three-dimensional femtoscopic analyses in terms of centrality classes and pair transverse momentum $k_{\rm T}$ ranges.

A decrease of source radii with increasing transverse momentum $k_\mathrm{T}$ (mass $m_\mathrm{T}$) and decreasing charged-particle multiplicity is observed. Within the hydrodynamic scenario, this trend is considered as a manifestation of collective effects in the matter created in heavy-ion collisions which weaken with decreasing multiplicity. The extracted one-dimensional radii and $\lambda$ parameters are in good agreement with the results of the previous ALICE analysis of identical charged-kaon correlations in Pb--Pb collisions at \mbox{$\sqrt{s_{\mathrm{NN}}} = 2.76$}~TeV. The three-dimensional experimental radii agree within uncertainties with the iHKM calculations, wherein the hydrodynamic phase is followed by the hadronic rescattering phase. However, the uncertainties of the experimental data do not allow for distinguishing between the two considered model scenarios corresponding to the two different equations of state. The model underestimates the experimental $R_\mathrm{out}$ for the 0--5\% centrality class by about 1.0--1.5~fm similarly to what was observed for $\pi$K correlations in Pb--Pb collisions at \mbox{$\sqrt{s_{\mathrm{NN}}} = 2.76$}~TeV and K$^\pm$K$^\pm$ correlations in p--Pb collisions at \mbox{$\sqrt{s_{\mathrm{NN}}} = 5.02$}~TeV, which must be addressed by further work on models of the system evolution in high-energy heavy-ion collisions. The extracted times of maximal emission for kaons decrease with decreasing charged-particle multiplicity, showing that the system created in more peripheral events evolves quickly.

The results obtained in the present analysis provide information about the dynamics of the matter created in heavy-ion collisions, which can be explained in terms of the hydrodynamic approach, including effects such as collective flow and hadron rescattering in the final state. They can provide critical constraints for tuning hydrodynamic models of heavy-ion collisions, specifically regarding the extracted homogeneity lengths and emission duration.

%%%%% acknowledgements - handled by EB chairs
\newenvironment{acknowledgement}{\relax}{\relax}
\begin{acknowledgement}
\section*{Acknowledgements}
% add specific acknowledgements here
% ...but please don't remove the line below: funding agencies
% will be acknowledged with a custom tex file handled by EB chairs after Collab Round 2
% Version: 2025-09-18

The ALICE Collaboration would like to thank all its engineers and technicians for their invaluable contributions to the construction of the experiment and the CERN accelerator teams for the outstanding performance of the LHC complex.
The ALICE Collaboration gratefully acknowledges the resources and support provided by all Grid centres and the Worldwide LHC Computing Grid (WLCG) collaboration.
The ALICE Collaboration acknowledges the following funding agencies for their support in building and running the ALICE detector:
A. I. Alikhanyan National Science Laboratory (Yerevan Physics Institute) Foundation (ANSL), State Committee of Science and World Federation of Scientists (WFS), Armenia;
Austrian Academy of Sciences, Austrian Science Fund (FWF): [M 2467-N36] and Nationalstiftung f\"{u}r Forschung, Technologie und Entwicklung, Austria;
Ministry of Communications and High Technologies, National Nuclear Research Center, Azerbaijan;
Rede Nacional de Física de Altas Energias (Renafae), Financiadora de Estudos e Projetos (Finep), Funda\c{c}\~{a}o de Amparo \`{a} Pesquisa do Estado de S\~{a}o Paulo (FAPESP) and The Sao Paulo Research Foundation  (FAPESP), Brazil;
Bulgarian Ministry of Education and Science, within the National Roadmap for Research Infrastructures 2020-2027 (object CERN), Bulgaria;
Ministry of Education of China (MOEC) , Ministry of Science \& Technology of China (MSTC) and National Natural Science Foundation of China (NSFC), China;
Ministry of Science and Education and Croatian Science Foundation, Croatia;
Centro de Aplicaciones Tecnol\'{o}gicas y Desarrollo Nuclear (CEADEN), Cubaenerg\'{\i}a, Cuba;
Ministry of Education, Youth and Sports of the Czech Republic, Czech Republic;
The Danish Council for Independent Research | Natural Sciences, the VILLUM FONDEN and Danish National Research Foundation (DNRF), Denmark;
Helsinki Institute of Physics (HIP), Finland;
Commissariat \`{a} l'Energie Atomique (CEA) and Institut National de Physique Nucl\'{e}aire et de Physique des Particules (IN2P3) and Centre National de la Recherche Scientifique (CNRS), France;
Bundesministerium f\"{u}r Forschung, Technologie und Raumfahrt (BMFTR) and GSI Helmholtzzentrum f\"{u}r Schwerionenforschung GmbH, Germany;
General Secretariat for Research and Technology, Ministry of Education, Research and Religions, Greece;
National Research, Development and Innovation Office, Hungary;
Department of Atomic Energy Government of India (DAE), Department of Science and Technology, Government of India (DST), University Grants Commission, Government of India (UGC) and Council of Scientific and Industrial Research (CSIR), India;
National Research and Innovation Agency - BRIN, Indonesia;
Istituto Nazionale di Fisica Nucleare (INFN), Italy;
Japanese Ministry of Education, Culture, Sports, Science and Technology (MEXT) and Japan Society for the Promotion of Science (JSPS) KAKENHI, Japan;
Consejo Nacional de Ciencia (CONACYT) y Tecnolog\'{i}a, through Fondo de Cooperaci\'{o}n Internacional en Ciencia y Tecnolog\'{i}a (FONCICYT) and Direcci\'{o}n General de Asuntos del Personal Academico (DGAPA), Mexico;
Nederlandse Organisatie voor Wetenschappelijk Onderzoek (NWO), Netherlands;
The Research Council of Norway, Norway;
Pontificia Universidad Cat\'{o}lica del Per\'{u}, Peru;
Ministry of Science and Higher Education, National Science Centre and WUT ID-UB, Poland;
Korea Institute of Science and Technology Information and National Research Foundation of Korea (NRF), Republic of Korea;
Ministry of Education and Scientific Research, Institute of Atomic Physics, Ministry of Research and Innovation and Institute of Atomic Physics and Universitatea Nationala de Stiinta si Tehnologie Politehnica Bucuresti, Romania;
Ministerstvo skolstva, vyskumu, vyvoja a mladeze SR, Slovakia;
National Research Foundation of South Africa, South Africa;
Swedish Research Council (VR) and Knut \& Alice Wallenberg Foundation (KAW), Sweden;
European Organization for Nuclear Research, Switzerland;
Suranaree University of Technology (SUT), National Science and Technology Development Agency (NSTDA) and National Science, Research and Innovation Fund (NSRF via PMU-B B05F650021), Thailand;
Turkish Energy, Nuclear and Mineral Research Agency (TENMAK), Turkey;
National Academy of  Sciences of Ukraine, Ukraine;
Science and Technology Facilities Council (STFC), United Kingdom;
National Science Foundation of the United States of America (NSF) and United States Department of Energy, Office of Nuclear Physics (DOE NP), United States of America.
In addition, individual groups or members have received support from:
Czech Science Foundation (grant no. 23-07499S), Czech Republic;
FORTE project, reg.\ no.\ CZ.02.01.01/00/22\_008/0004632, Czech Republic, co-funded by the European Union, Czech Republic;
European Research Council (grant no. 950692), European Union;
Deutsche Forschungs Gemeinschaft (DFG, German Research Foundation) ``Neutrinos and Dark Matter in Astro- and Particle Physics'' (grant no. SFB 1258), Germany;
CONVECS project, CUP C97H23001700002 FESR 2021-2027 program, Italy;
FAIR - Future Artificial Intelligence Research, funded by the NextGenerationEU program (Italy).

\end{acknowledgement}

%%%%%%%% Bibliography
\bibliographystyle{utphys}   % Remember we use title in the biblio
\bibliography{bibliography}
%\input {bibliography.tex}

%%%%%%%%%%%%%%%%%%%%%%%%%%%%%%%%
% Appendices: yours (if any) + authorlist
%%%%%%%%%%%%%%%%%%%%%%%%%%%%%%%%
\newpage
\appendix

%
%\input{} % put your appendices here (if any)
%

%%%%% Authorlist - please do not touch: handled by EB chairs
\section{The ALICE Collaboration}
\label{app:collab}
% ALICE Collaboration author list for 2025-09-18
\begin{flushleft} 
\small

I.J.~Abualrob\,\orcidlink{0009-0005-3519-5631}\,$^{\rm 114}$, 
S.~Acharya\,\orcidlink{0000-0002-9213-5329}\,$^{\rm 50}$, 
G.~Aglieri Rinella\,\orcidlink{0000-0002-9611-3696}\,$^{\rm 32}$, 
L.~Aglietta\,\orcidlink{0009-0003-0763-6802}\,$^{\rm 24}$, 
N.~Agrawal\,\orcidlink{0000-0003-0348-9836}\,$^{\rm 25}$, 
Z.~Ahammed\,\orcidlink{0000-0001-5241-7412}\,$^{\rm 134}$, 
S.~Ahmad\,\orcidlink{0000-0003-0497-5705}\,$^{\rm 15}$, 
I.~Ahuja\,\orcidlink{0000-0002-4417-1392}\,$^{\rm 36}$, 
Z.~Akbar$^{\rm 81}$, 
A.~Akindinov\,\orcidlink{0000-0002-7388-3022}\,$^{\rm 140}$
% ...

\section*{Affiliation Notes}

$^{\rm I}$ Deceased\\
$^{\rm II}$ Also at: Max-Planck-Institut fur Physik, Munich, Germany\\
$^{\rm III}$ Also at: Czech Technical University in Prague, Prague, Czech Republic\\
$^{\rm IV}$ Also at: Instituto de Fisica da Universidade de Sao Paulo\\
$^{\rm V}$ Also at: Dipartimento DET del Politecnico di Torino, Turin, Italy\\
$^{\rm VI}$ Also at: University College of Dublin, Dublin, Ireland\\
$^{\rm VII}$ Also at: Department of Applied Physics, Aligarh Muslim University, Aligarh, India\\
$^{\rm VIII}$ Also at: Institute of Theoretical Physics, University of Wroclaw, Wroclaw, Poland\\
$^{\rm IX}$ Also at: Facultad de Ciencias, Universidad Nacional Aut\'{o}noma de M\'{e}xico, Mexico City, Mexico\\
$^{\rm X}$ Also at: Lovely Professional University, Punjab, India\\

\section*{Collaboration Institutes}

$^{1}$ A.I. Alikhanyan National Science Laboratory (Yerevan Physics Institute) Foundation, Yerevan, Armenia\\
$^{2}$ AGH University of Krakow, Cracow, Poland\\
$^{3}$ Bogolyubov Institute for Theoretical Physics, National Academy of Sciences of Ukraine, Kyiv, Ukraine\\
$^{4}$ Bose Institute, Department of Physics  and Centre for Astroparticle Physics and Space Science (CAPSS), Kolkata, India\\
$^{5}$ California Polytechnic State University, San Luis Obispo, California, United States\\
$^{6}$ Central China Normal University, Wuhan, China\\
$^{7}$ Centro de Aplicaciones Tecnol\'{o}gicas y Desarrollo Nuclear (CEADEN), Havana, Cuba\\
$^{8}$ Centro de Investigaci\'{o}n y de Estudios Avanzados (CINVESTAV), Mexico City and M\'{e}rida, Mexico\\
$^{9}$ Chicago State University, Chicago, Illinois, United States\\
$^{10}$ China Nuclear Data Center, China Institute of Atomic Energy, Beijing, China\\
$^{11}$ China University of Geosciences, Wuhan, China\\
$^{12}$ Chungbuk National University, Cheongju, Republic of Korea\\
$^{13}$ Comenius University Bratislava, Faculty of Mathematics, Physics and Informatics, Bratislava, Slovak Republic\\
$^{14}$ Creighton University, Omaha, Nebraska, United States\\
$^{15}$ Department of Physics, Aligarh Muslim University, Aligarh, India\\
$^{16}$ Department of Physics, Pusan National University, Pusan, Republic of Korea\\
$^{17}$ Department of Physics, Sejong University, Seoul, Republic of Korea\\
$^{18}$ Department of Physics, University of California, Berkeley, California, United States\\
$^{19}$ Department of Physics, University of Oslo, Oslo, Norway\\
$^{20}$ Department of Physics and Technology, University of Bergen, Bergen, Norway\\
$^{21}$ Dipartimento di Fisica, Universit\`{a} di Pavia, Pavia, Italy\\
$^{22}$ Dipartimento di Fisica dell'Universit\`{a} and Sezione INFN, Cagliari, Italy\\
$^{23}$ Dipartimento di Fisica dell'Universit\`{a} and Sezione INFN, Trieste, Italy\\
$^{24}$ Dipartimento di Fisica dell'Universit\`{a} and Sezione INFN, Turin, Italy\\
$^{25}$ Dipartimento di Fisica e Astronomia dell'Universit\`{a} and Sezione INFN, Bologna, Italy\\
$^{26}$ Dipartimento di Fisica e Astronomia dell'Universit\`{a} and Sezione INFN, Catania, Italy\\
$^{27}$ Dipartimento di Fisica e Astronomia dell'Universit\`{a} and Sezione INFN, Padova, Italy\\
$^{28}$ Dipartimento di Fisica `E.R.~Caianiello' dell'Universit\`{a} and Gruppo Collegato INFN, Salerno, Italy\\
$^{29}$ Dipartimento DISAT del Politecnico and Sezione INFN, Turin, Italy\\
$^{30}$ Dipartimento di Scienze MIFT, Universit\`{a} di Messina, Messina, Italy\\
$^{31}$ Dipartimento Interateneo di Fisica `M.~Merlin' and Sezione INFN, Bari, Italy\\
$^{32}$ European Organization for Nuclear Research (CERN), Geneva, Switzerland\\
$^{33}$ Faculty of Electrical Engineering, Mechanical Engineering and Naval Architecture, University of Split, Split, Croatia\\
$^{34}$ Faculty of Nuclear Sciences and Physical Engineering, Czech Technical University in Prague, Prague, Czech Republic\\
$^{35}$ Faculty of Physics, Sofia University, Sofia, Bulgaria\\
$^{36}$ Faculty of Science, P.J.~\v{S}af\'{a}rik University, Ko\v{s}ice, Slovak Republic\\
$^{37}$ Faculty of Technology, Environmental and Social Sciences, Bergen, Norway\\
$^{38}$ Frankfurt Institute for Advanced Studies, Johann Wolfgang Goethe-Universit\"{a}t Frankfurt, Frankfurt, Germany\\
$^{39}$ Fudan University, Shanghai, China\\
$^{40}$ Gangneung-Wonju National University, Gangneung, Republic of Korea\\
$^{41}$ Gauhati University, Department of Physics, Guwahati, India\\
$^{42}$ Helmholtz-Institut f\"{u}r Strahlen- und Kernphysik, Rheinische Friedrich-Wilhelms-Universit\"{a}t Bonn, Bonn, Germany\\
$^{43}$ Helsinki Institute of Physics (HIP), Helsinki, Finland\\
$^{44}$ High Energy Physics Group,  Universidad Aut\'{o}noma de Puebla, Puebla, Mexico\\
$^{45}$ Horia Hulubei National Institute of Physics and Nuclear Engineering, Bucharest, Romania\\
$^{46}$ HUN-REN Wigner Research Centre for Physics, Budapest, Hungary\\
$^{47}$ Indian Institute of Technology Bombay (IIT), Mumbai, India\\
$^{48}$ Indian Institute of Technology Indore, Indore, India\\
$^{49}$ INFN, Laboratori Nazionali di Frascati, Frascati, Italy\\
$^{50}$ INFN, Sezione di Bari, Bari, Italy\\
$^{51}$ INFN, Sezione di Bologna, Bologna, Italy\\
$^{52}$ INFN, Sezione di Cagliari, Cagliari, Italy\\
$^{53}$ INFN, Sezione di Catania, Catania, Italy\\
$^{54}$ INFN, Sezione di Padova, Padova, Italy\\
$^{55}$ INFN, Sezione di Pavia, Pavia, Italy\\
$^{56}$ INFN, Sezione di Torino, Turin, Italy\\
$^{57}$ INFN, Sezione di Trieste, Trieste, Italy\\
$^{58}$ Inha University, Incheon, Republic of Korea\\
$^{59}$ Institute for Gravitational and Subatomic Physics (GRASP), Utrecht University/Nikhef, Utrecht, Netherlands\\
$^{60}$ Institute of Experimental Physics, Slovak Academy of Sciences, Ko\v{s}ice, Slovak Republic\\
$^{61}$ Institute of Physics, Homi Bhabha National Institute, Bhubaneswar, India\\
$^{62}$ Institute of Physics of the Czech Academy of Sciences, Prague, Czech Republic\\
$^{63}$ Institute of Space Science (ISS), Bucharest, Romania\\
$^{64}$ Institut f\"{u}r Kernphysik, Johann Wolfgang Goethe-Universit\"{a}t Frankfurt, Frankfurt, Germany\\
$^{65}$ Instituto de Ciencias Nucleares, Universidad Nacional Aut\'{o}noma de M\'{e}xico, Mexico City, Mexico\\
$^{66}$ Instituto de F\'{i}sica, Universidade Federal do Rio Grande do Sul (UFRGS), Porto Alegre, Brazil\\
$^{67}$ Instituto de F\'{\i}sica, Universidad Nacional Aut\'{o}noma de M\'{e}xico, Mexico City, Mexico\\
$^{68}$ iThemba LABS, National Research Foundation, Somerset West, South Africa\\
$^{69}$ Jeonbuk National University, Jeonju, Republic of Korea\\
$^{70}$ Korea Institute of Science and Technology Information, Daejeon, Republic of Korea\\
$^{71}$ Laboratoire de Physique Subatomique et de Cosmologie, Universit\'{e} Grenoble-Alpes, CNRS-IN2P3, Grenoble, France\\
$^{72}$ Lawrence Berkeley National Laboratory, Berkeley, California, United States\\
$^{73}$ Lund University Department of Physics, Division of Particle Physics, Lund, Sweden\\
$^{74}$ Marietta Blau Institute, Vienna, Austria\\
$^{75}$ Nagasaki Institute of Applied Science, Nagasaki, Japan\\
$^{76}$ Nara Women{'}s University (NWU), Nara, Japan\\
$^{77}$ National and Kapodistrian University of Athens, School of Science, Department of Physics , Athens, Greece\\
$^{78}$ National Centre for Nuclear Research, Warsaw, Poland\\
$^{79}$ National Institute of Science Education and Research, Homi Bhabha National Institute, Jatni, India\\
$^{80}$ National Nuclear Research Center, Baku, Azerbaijan\\
$^{81}$ National Research and Innovation Agency - BRIN, Jakarta, Indonesia\\
$^{82}$ Niels Bohr Institute, University of Copenhagen, Copenhagen, Denmark\\
$^{83}$ Nikhef, National institute for subatomic physics, Amsterdam, Netherlands\\
$^{84}$ Nuclear Physics Group, STFC Daresbury Laboratory, Daresbury, United Kingdom\\
$^{85}$ Nuclear Physics Institute of the Czech Academy of Sciences, Husinec-\v{R}e\v{z}, Czech Republic\\
$^{86}$ Oak Ridge National Laboratory, Oak Ridge, Tennessee, United States\\
$^{87}$ Ohio State University, Columbus, Ohio, United States\\
$^{88}$ Physics department, Faculty of science, University of Zagreb, Zagreb, Croatia\\
$^{89}$ Physics Department, Panjab University, Chandigarh, India\\
$^{90}$ Physics Department, University of Jammu, Jammu, India\\
$^{91}$ Physics Program and International Institute for Sustainability with Knotted Chiral Meta Matter (WPI-SKCM$^{2}$), Hiroshima University, Hiroshima, Japan\\
$^{92}$ Physikalisches Institut, Eberhard-Karls-Universit\"{a}t T\"{u}bingen, T\"{u}bingen, Germany\\
$^{93}$ Physikalisches Institut, Ruprecht-Karls-Universit\"{a}t Heidelberg, Heidelberg, Germany\\
$^{94}$ Physik Department, Technische Universit\"{a}t M\"{u}nchen, Munich, Germany\\
$^{95}$ Politecnico di Bari and Sezione INFN, Bari, Italy\\
$^{96}$ Research Division and ExtreMe Matter Institute EMMI, GSI Helmholtzzentrum f\"ur Schwerionenforschung GmbH, Darmstadt, Germany\\
$^{97}$ Saga University, Saga, Japan\\
$^{98}$ Saha Institute of Nuclear Physics, Homi Bhabha National Institute, Kolkata, India\\
$^{99}$ School of Physics and Astronomy, University of Birmingham, Birmingham, United Kingdom\\
$^{100}$ Secci\'{o}n F\'{\i}sica, Departamento de Ciencias, Pontificia Universidad Cat\'{o}lica del Per\'{u}, Lima, Peru\\
$^{101}$ SUBATECH, IMT Atlantique, Nantes Universit\'{e}, CNRS-IN2P3, Nantes, France\\
$^{102}$ Sungkyunkwan University, Suwon City, Republic of Korea\\
$^{103}$ Suranaree University of Technology, Nakhon Ratchasima, Thailand\\
$^{104}$ Technical University of Ko\v{s}ice, Ko\v{s}ice, Slovak Republic\\
$^{105}$ The Henryk Niewodniczanski Institute of Nuclear Physics, Polish Academy of Sciences, Cracow, Poland\\
$^{106}$ The University of Texas at Austin, Austin, Texas, United States\\
$^{107}$ Universidad Aut\'{o}noma de Sinaloa, Culiac\'{a}n, Mexico\\
$^{108}$ Universidade de S\~{a}o Paulo (USP), S\~{a}o Paulo, Brazil\\
$^{109}$ Universidade Estadual de Campinas (UNICAMP), Campinas, Brazil\\
$^{110}$ Universidade Federal do ABC, Santo Andre, Brazil\\
$^{111}$ Universitatea Nationala de Stiinta si Tehnologie Politehnica Bucuresti, Bucharest, Romania\\
$^{112}$ University of Cape Town, Cape Town, South Africa\\
$^{113}$ University of Derby, Derby, United Kingdom\\
$^{114}$ University of Houston, Houston, Texas, United States\\
$^{115}$ University of Jyv\"{a}skyl\"{a}, Jyv\"{a}skyl\"{a}, Finland\\
$^{116}$ University of Kansas, Lawrence, Kansas, United States\\
$^{117}$ University of Liverpool, Liverpool, United Kingdom\\
$^{118}$ University of Science and Technology of China, Hefei, China\\
$^{119}$ University of Silesia in Katowice, Katowice, Poland\\
$^{120}$ University of South-Eastern Norway, Kongsberg, Norway\\
$^{121}$ University of Tennessee, Knoxville, Tennessee, United States\\
$^{122}$ University of the Witwatersrand, Johannesburg, South Africa\\
$^{123}$ University of Tokyo, Tokyo, Japan\\
$^{124}$ University of Tsukuba, Tsukuba, Japan\\
$^{125}$ Universit\"{a}t M\"{u}nster, Institut f\"{u}r Kernphysik, M\"{u}nster, Germany\\
$^{126}$ Universit\'{e} Clermont Auvergne, CNRS/IN2P3, LPC, Clermont-Ferrand, France\\
$^{127}$ Universit\'{e} de Lyon, CNRS/IN2P3, Institut de Physique des 2 Infinis de Lyon, Lyon, France\\
$^{128}$ Universit\'{e} de Strasbourg, CNRS, IPHC UMR 7178, F-67000 Strasbourg, France, Strasbourg, France\\
$^{129}$ Universit\'{e} Paris-Saclay, Centre d'Etudes de Saclay (CEA), IRFU, D\'{e}partment de Physique Nucl\'{e}aire (DPhN), Saclay, France\\
$^{130}$ Universit\'{e}  Paris-Saclay, CNRS/IN2P3, IJCLab, Orsay, France\\
$^{131}$ Universit\`{a} degli Studi di Foggia, Foggia, Italy\\
$^{132}$ Universit\`{a} del Piemonte Orientale, Vercelli, Italy\\
$^{133}$ Universit\`{a} di Brescia, Brescia, Italy\\
$^{134}$ Variable Energy Cyclotron Centre, Homi Bhabha National Institute, Kolkata, India\\
$^{135}$ Warsaw University of Technology, Warsaw, Poland\\
$^{136}$ Wayne State University, Detroit, Michigan, United States\\
$^{137}$ Yale University, New Haven, Connecticut, United States\\
$^{138}$ Yildiz Technical University, Istanbul, Turkey\\
$^{139}$ Yonsei University, Seoul, Republic of Korea\\
$^{140}$ Affiliated with an institute formerly covered by a cooperation agreement with CERN\\
$^{141}$ Affiliated with an international laboratory covered by a cooperation agreement with CERN.\\

\end{flushleft} 

\end{document}